\documentclass[pra,aps,showpacs,groupedaddress,superscriptaddress,twocolumn,toc=flat,nofootinbib]{revtex4-1}
\usepackage[colorlinks=true,linkcolor=blue,urlcolor=blue,citecolor=blue]{hyperref}

\usepackage[utf8x]{inputenc}
\usepackage{color}
\usepackage{bbm} 

\usepackage{amsfonts,amsmath,amssymb,stmaryrd}

\usepackage{graphicx}
\usepackage{subfigure}  
\usepackage{bbm} 
\usepackage{hyperref}
\usepackage{epsfig}
\usepackage{mathrsfs}
\usepackage{verbatim}
\usepackage{centernot}
\usepackage{ulem}
\usepackage{longtable}
\usepackage{nicefrac}
\usepackage[framemethod=tikz]{mdframed}
 
\renewcommand{\l}{\left(}
\renewcommand{\r}{\right)}

\newcommand{\Zt}{$\mathbb{Z}_2$~}

\newcommand{\bra}[1]{\langle#1|}
\newcommand{\ket}[1]{|#1\rangle}

\renewcommand{\ij}{{\langle i, j \rangle}}

\renewcommand{\H}{\hat{\mathcal{H}}}

\renewcommand{\c}{\hat{c}}
\renewcommand{\a}{\hat{a}}

\newcommand{\cd}{\hat{c}^\dagger}

\newcommand{\ad}{\hat{a}^\dagger}

\newcommand{\n}{\hat{n}}

\newcommand{\hc}{\text{h.c.}}

\usepackage{array}

\usepackage{cancel,ifthen}
\newcommand{\cmnt}[2][NoInPuT]{\ifthenelse{\equal{#1}{NoInPuT}}{}{{\color{red}\sout{#1}}} {\color{blue} #2}}

\usepackage{bm}	

\LTcapwidth=\textwidth

\begin{document}
\normalem	

\title{Confinement and Mott transitions of dynamical charges in 1D lattice gauge theories}

\author{Matja\v{z} Kebri\v{c}}
\affiliation{Department of Physics and Arnold Sommerfeld Center for Theoretical Physics (ASC), Ludwig-Maximilians-Universit\"at M\"unchen, Theresienstr. 37, M\"unchen D-80333, Germany}
\affiliation{Munich Center for Quantum Science and Technology (MCQST), Schellingstr. 4, D-80799 M\"unchen, Germany}

\author{Luca Barbiero}
\affiliation{ICFO - Institut de Ciències Fotòniques, The Barcelona Institute of Science and Technology, 08860 Castelldefels (Barcelona), Spain.}
\affiliation{Center for Nonlinear Phenomena and Complex Systems, Université Libre de Bruxelles, CP 231, Campus Plaine, B-1050 Brussels, Belgium.}

\author{Christian Reinmoser}
\affiliation{Department of Physics and Arnold Sommerfeld Center for Theoretical Physics (ASC), Ludwig-Maximilians-Universit\"at M\"unchen, Theresienstr. 37, M\"unchen D-80333, Germany}
\affiliation{Munich Center for Quantum Science and Technology (MCQST), Schellingstr. 4, D-80799 M\"unchen, Germany}

\author{Ulrich Schollw\"ock}
\affiliation{Department of Physics and Arnold Sommerfeld Center for Theoretical Physics (ASC), Ludwig-Maximilians-Universit\"at M\"unchen, Theresienstr. 37, M\"unchen D-80333, Germany}
\affiliation{Munich Center for Quantum Science and Technology (MCQST), Schellingstr. 4, D-80799 M\"unchen, Germany}

\author{Fabian Grusdt}
\email[Corresponding author email: ]{fabian.grusdt@physik.uni-muenchen.de}
\affiliation{Department of Physics and Arnold Sommerfeld Center for Theoretical Physics (ASC), Ludwig-Maximilians-Universit\"at M\"unchen, Theresienstr. 37, M\"unchen D-80333, Germany}
\affiliation{Munich Center for Quantum Science and Technology (MCQST), Schellingstr. 4, D-80799 M\"unchen, Germany}

\pacs{}

\date{\today}

\begin{abstract}
Confinement is an ubiquitous phenomenon when matter couples to gauge fields, which manifests itself in a linear string potential between two static charges. Although gauge fields can be integrated out in one dimension, they can mediate non-local interactions which in turn influence the paradigmatic Luttinger liquid properties. However, when the charges become dynamical and their densities finite, understanding confinement becomes challenging. Here we show that confinement in 1D $\mathbb{Z}_2$ lattice gauge theories, with dynamical matter fields and arbitrary densities, is related to translational symmetry breaking in a non-local basis. The exact transformation to this string-length basis leads us to an exact mapping of Luttinger parameters reminiscent of a Luther-Emery re-scaling. We include the effects of local, but beyond contact, interactions between the matter particles, and show that confined mesons can form a Mott-insulating state when the deconfined charges cannot. While the transition to the Mott state cannot be detected in the Green's function of the charges, we show that the metallic state is characterized by hidden off-diagonal quasi-long range order. Our predictions provide new insights to the physics of confinement of dynamical charges, and can be experimentally addressed in Rydberg-dressed quantum gases in optical lattices. 
\end{abstract}

\maketitle

\emph{Introduction.--}
Lattice gauge theories (LGTs), originally introduced to get insights about non-perturbative regimes in particle physics \cite{Kogut1979, Wilson1974}, have become a powerful tool to tackle many-body problems in condensed matter systems \cite{Wen2004, Levin2005, Lee2006}. These theories turn out to be particularly rich and interesting when the matter is coupled to dynamical gauge fields: For example, in some cases the confinement-deconfinement transition \cite{Kogut1979} can be associated with the appearance of topological phases with non-Abelian anyons and charge fractionalization \cite{Kitaev2003}. On the other hand, when the matter acquires its own quantum dynamics the confinement problem is poorly understood and, in this regime, a general physical description of the phenomenon is still lacking. Furthermore, the high level of complexity of LGTs makes theoretical studies based on standard numerical methods \cite{Troyer2005, Alford2008, Magnifico2020, Kuno2017} very challenging.

\begin{figure}[t!]
\centering
\epsfig{file=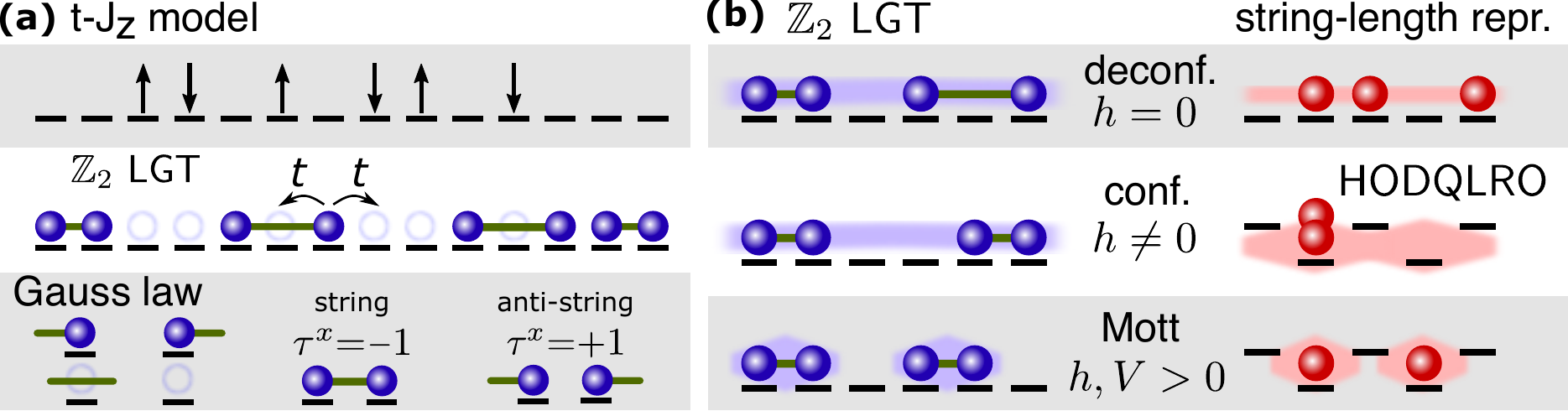, width=0.48\textwidth}
\caption{(a) The 1D $t-J_{z}$ model (top) maps exactly to 1D $\mathbb{Z}_{2}$ LGTs (middle); blue full spheres correspond to hard-core bosons or fermions and the empty blue circles denote holes. Pairs of particles are connected with green lines, which correspond to \Zt electric strings, according to the configurations allowed by the Gauss law (bottom). (b) Comparison between the deconfined, confined and Mott states, both in the original and in the corresponding string-length representation. As indicated, the confining phases are characterized by a broken translational symmetry in the string-length basis.} 
\label{figCombinedLGT}
\end{figure}

At the same time, due to their impressive level of control and accuracy, ultracold atomic systems are establishing themselves as a fundamental platform where LGT models can be systematically studied \cite{Wiese2013, Zohar2013, Zohar2015, Bender2018, Dalmonte2016, Martinez2016, Schweizer2019, Goerg2019, Mil2020, Yang2020}. In this context LGTs with an Ising gauge group, i.e. \Zt LGTs \cite{Zohar2017, Barbiero2019,Homeier2020}, are particularly meaningful to explore, allowing for instance to study their connections to strongly correlated electronic systems \cite{Sedgewick2002, Demler2002, Kaul2007, Sachdev2016} including high-$T_c$ superconductivity \cite{Senthil2000, Lee2007}. Recent theoretical studies of two dimensional \Zt LGTs with matter-gauge coupling have revealed a wealth of intriguing properties \cite{Gazit2017, Borla2021, Borla2020a, Seifert2020}. Experimentally, a first instance of a \Zt LGT with dynamical matter has recently been realized in a mixture of ultracold bosons in a double well potential \cite{Schweizer2019} by means of a Floquet scheme \cite{Barbiero2019}. Using an extension of this Floquet scheme \cite{Barbiero2019,Goerg2019}, or coupling superconducting qubits \cite{Zohar2017,Homeier2020}, allows to study \Zt LGTs with dynamical matter in extended geometries and higher dimensions, thus paving the way towards a deeper understanding of such models. Moreover, as it will be discussed below, in one dimension (1D) the direct implementation of Hamiltonians with encoded gauge degrees of freedom \cite{Grusdt2020, Cuadra2020} can also be employed to explore \Zt LGTs, see Fig.~\ref{figCombinedLGT} (a).

In this letter, we solve the confinement problem in a class of 1D \Zt LGTs with dynamical charges \cite{Borla2020} at arbitrary densities. This is achieved by representing the \Zt LGT model in the non-local basis of string lengths, where we prove that confinement is equivalent to a broken translational symmetry. Our argument applies for a larger class of 1D LGTs. 
We also study the Mott transition of \Zt charges, which defies conventional wisdom for at least two reasons. 

First we show that an exponentially decaying \Zt invariant Green's function no longer provides a unique signature of the Mott state. Instead, we show that the confined Luttinger liquid \cite{Borla2020} is characterized by hidden off-diagonal quasi-long range order (HODQLRO) in the string-length basis. This quasi-condensate of string excitations is destroyed at the Mott transition, see Fig.~\ref{figCombinedLGT} (b). More formally, we derive a Luther-Emery like relation between the Luttinger parameters in the original \Zt LGT and the effective model in the string-length basis.

Second we show that the Mott insulator occurring at the specific filling $n^a=2/3$ is stabilized by a combination of the attractive confining potential and a nearest-neighbor (NN) repulsion. If either of those terms is absent, a gapless liquid is obtained; on the other hand when both are sizable our numerical simulations, based on density-matrix-renormalization-group (DMRG) algorithm  \cite{White1992, Schollwoeck2011, hubig:_syten_toolk, hubig17:_symmet_protec_tensor_networ}, yield a significant charge gap. Our predictions can be tested in Rydberg-dressed atomic gases in optical lattices, where site-resolved quantum projective measurements provide direct access to the non-local string-length basis.

\emph{Model.--}
We consider a 1D \Zt LGT Hamiltonian where $N$ hard-core bosons in a lattice with $L$ sites are coupled to \Zt gauge fields,
\begin{equation}
 	\H = - t \sum_{\ij} \l \ad_i \hat{\tau}^z_{\ij} \a_j + \hc \r - h \sum_{\ij} \hat{\tau}^x_{\ij} + V \sum_\ij \hat{n}_i \hat{n}_j .
	\label{eqDefModel}
\end{equation}
Here $\ad_i$ denotes the hard-core bosonic creation operator, $t$ describes NN tunneling processes mediated by the \Zt gauge field $\hat{\tau}^z_{\ij}$ defined on the links $\ij$ between NN sites, and $V>0$ represents a NN repulsion between bosons. The gauge-invariant \Zt electric-field term $\hat{\tau}^x_{\ij}$ with strength $h$ introduces quantum fluctuations of $\hat{\tau}^z_{\ij}$. The physics remains unchanged if $\hat{a}$'s are replaced by fermionic operators $\hat{c}$, as can be shown by a Jordan-Wigner transformation. 

The \Zt electric-field is subject to a Gauss law which ensures that the former changes sign across a particle \cite{Borla2020}; i.e. pairs of particles are connected by \Zt electric-fields of the same sign, which we denote as \Zt electric strings and anti-strings, see Fig.~\ref{figCombinedLGT} (a). For concreteness, we assume open boundary conditions with $\tau^x_{\langle 0, 1 \rangle} = 1$ (no \Zt electric string entering from the left). The corresponding \Zt gauge group is defined by the operator
\begin{equation}
 	\hat{G}_{i} =  \hat{\tau}^{x}_{\left \langle i-1, i \right \rangle} \hat{\tau}^{x}_{\left \langle i, i+1 \right \rangle} (-1)^{\hat{n}_{i}},
	\label{eqGaussLaw}
\end{equation}
which commutes with the Hamiltonian $[\H, \hat{G}_{i}] = 0$ and itself $[\hat{G}_{i}, \hat{G}_{j}] = 0$. As a consequence, the effective Hilbert space of eq. (\ref{eqDefModel}) is split into different sectors $\hat{G}_{i} = \pm 1$ \cite{Borla2020, Prosko2017}. The Gauss law we choose corresponds to the sector where $\hat{G}_{i} = 1, \forall i$, see Fig.~\ref{figCombinedLGT}.

\emph{Implementation.--}
In order to implement the LGT Hamiltonian Eq.~\eqref{eqDefModel}, we propose a Rydberg dressing scheme in a spin-dependent super-lattice potential with period $2a$, $a$ being the lattice spacing. We require the following potential,
\begin{equation}
	V_{\sigma, j} = \left (-1 \right)^{\sigma} \frac{\omega_{0}}{2} -  \left( -1 \right)^{\sigma} \frac{\delta}{2} \left( 1- \left(- 1 \right)^{j} \right),
	\label{eqPotential}
\end{equation}
where the first term describes the splitting $\omega_{0}$ between the two spin states. The second term realizes a staggered magnetic Zeeman field and can be realized by an anti-magic superlattice, e.g. using Ytterbium atoms \cite{Gerbier2010, Yi2008, Yang2017}. We propose to realize the required NN Ising interactions by dressing the spin states independently by two Rydberg dressing lasers $\Omega_{\uparrow}$ and $\Omega_{\downarrow}$ -- see \cite{SMPhysRev} for details.

This scheme gives an effective $t-J_z$ Hamiltonian \cite{Batista2000, Montorsi2020} which is best written in a rotating frame \cite{SMPhysRev},
\begin{multline}
	\H_{t-J_{z}} = \H _{t} +\sum_{\langle i, j \rangle} \bigg ( J_{\uparrow \uparrow} \hat{n}_{i}^{\uparrow} \hat{n}_{j}^{\uparrow} + J_{\uparrow \downarrow} \big ( \hat{n}_{i}^{\uparrow} \hat{n}_{j}^{\downarrow} + \hat{n}_{i}^{\downarrow} \hat{n}_{j}^{\uparrow}  \big ) \\ + J_{\downarrow \downarrow} \hat{n}_{i}^{\downarrow} \hat{n}_{j}^{\downarrow}  \bigg ) + \delta \sum_{j} (-1)^{j} \hat{S}^{z}_{j},
	\label{eqRydbergEffectiveNN}
\end{multline}
where $\H _{t} = -t \sum_{\langle i, j \rangle} \sum_\sigma \hat{P} (\hat{a}_{i,\sigma}^{\dagger} \hat{a}_{j,\sigma} + \hc ) \hat{P}$ is the hopping term, projected into the subspace without double occupancies. This model maps to the LGT model Eq.~\eqref{eqDefModel} by introducing a constraint on our Hilbert space where opposite spins appear in alternating fashion, leading to the \Zt Gauss law $\hat{G}_{i} = +1$ (for details see \cite{SMPhysRev}).

\emph{Confinement in \Zt LGTs.--}
In order to observe the confined and deconfined phases the \Zt gauge-invariant equal-time Green's function is considered
\begin{equation}
 g^{(1)}(i-j) = \langle \ad_i ~ \prod_{i \leq \ell \leq j} \hat{\tau}^z_\ell  ~ \a_j \rangle.
 \label{eqZtGreen}
\end{equation}
An algebraic decay of the correlator Eq.~\eqref{eqZtGreen} signals a deconfined phase where the charges can move around freely. An exponential decay, on the other hand, signals a confined phase where the particles are bound in pairs \cite{Borla2020}.

The Gauss law, $\hat{\tau}^{x}_{\left \langle i-1, i \right \rangle} = \hat{\tau}^{x}_{\left \langle i, i+1 \right \rangle} (-1)^{\hat{n}_{i}}$, can be successively applied to express the \Zt electric field as \cite{Borla2020}
\begin{equation}
 	\hat{\tau}^x_{\langle i,i+1 \rangle}  = \cos \bigl( \pi \sum_{j<i} \hat{n}_j \bigr),
	\label{eqTauXByN}
\end{equation}
which leads to a non-local term $- h \sum_{i} \cos \bigl( \pi \sum_{j<i} \hat{n}_j \bigr) $ in the Hamiltonian. By rewriting the density as $\n_{j} \rightarrow n(x) = n^{a} - \partial_{x} \phi(x) / \pi$, where $n^a$ is the average density, the \Zt electric field term after bosonization becomes
\begin{equation}
	-h \int  \mathrm{d} x \cos \bigl( \pi n^{a} x - \phi(x) \bigr).
	\label{eqTauXByNCont}
\end{equation}
Such oscillatory integrals should vanish, which means that the term is RG irrelevant \cite{Borla2020}. By this appealing but naive argument, the field term would be negligible and the system should behave like free fermions \cite{Prosko2017}. Hence, from this standard bosonization argument, one expects the correlator Eq.~\eqref{eqZtGreen} to have an algebraic decay $ g^{(1)}(d) \simeq |d|^{- \alpha}$ for non-zero values of $h \neq 0$ \cite{Giamarchi2004}. However, as shown in \cite{Borla2020} the decay is exponential -- and the charges confined -- for any $h \neq 0$. We attribute this failure of naive bosonization arguments to the non-local nature of the field term, Eq.~\eqref{eqTauXByN}, emphasized above.

Next we provide a general argument under which conditions the model in Eq.~\eqref{eqDefModel} is confining. To this end, we introduce a new non-local basis in which the Hamiltonian becomes local, meaning that conventional bosonization arguments can be safely applied.

\emph{String-length representation.--}
So far we represented basis states in our model by hard-core boson occupation numbers $n_j = 0,1$ and the \Zt electric strings $\tau^x_{\ij} = \pm 1$; as shown in Eq.~\eqref{eqTauXByN} the latter can be expressed by the former. Now we introduce new bosonic occupation numbers $\ell_1,...,\ell_{N+1} \geq 0$ to label our basis states, where $N = \sum_j n_j$ is the total conserved boson number. 

If $x_1,...,x_N$ denote the positions ($x_j=1,...,L$) of hard-core bosons, we define 
\begin{equation}
 	\ell_1 = x_1 - 1, \quad \ell_n = x_{n} - x_{n-1} - 1, \quad \ell_{N+1} = L - x_{N}.
	\label{eqDefStringBasis}
\end{equation}
This allows us to identify the corresponding Fock configuration $\ket{n_1,...,n_L}$ with a bosonic Fock configuration:
\begin{equation}
 	\ket{n_1,...,n_L} = \ket{\ell_1,...,\ell_{N+1}} \equiv \prod_{n=1}^{N+1} \frac{ ( \hat{\Psi}^\dagger_n )^{\ell_n} }{ \sqrt{\ell_n !} } \ket{0}.
\end{equation}
In the last step we introduced bosonic operators $\hat{\Psi}^\dagger_n$ acting on the string-length vacuum $\ket{0}$. 

Physically, the integers $\ell_n \in \mathbb{Z}_{\geq 0}$ describe the length of the \Zt (anti-) strings connecting pairs of consecutive \Zt charges, up to a shift of one: the shortest possible string connecting charges on NN sites is counted as having no excitation, $\ell=0$. The total number of string excitations, $\tilde{N} \equiv \sum_{\ell=1}^{N+1} \ell_n = L - N$, is conserved.

In the new string-length basis, we can express the \Zt LGT Hamiltonian as
\begin{multline}
 \H = - t \sum_{\langle m,n \rangle} \bigl( \hat{\rho}^{-1/2}_m  \hat{\Psi}^\dagger_m \hat{\Psi}_n  \hat{\rho}^{-1/2}_n + \hc \bigr) \\
  -  h \sum_n  (-1)^n \hat{\rho}_n + V \sum_n \delta_{\hat{\rho}_n,0}.
  \label{eqHtransfrm}
\end{multline}
Here $\delta_{a,b}$ denotes the Kronecker delta and $\hat{\rho}_n = \hat{\Psi}^\dagger_n \hat{\Psi}_n$ is the string-length density operator. The transformed Hamiltonian \eqref{eqHtransfrm} is purely local. It is defined on a lattice of size $\tilde{L} = N+1$ with $\tilde{N}$ excitations; i.e. the average boson density in this model is given by
\begin{equation}
 \rho^\Psi =  \frac{\tilde{N} }{ \tilde{L} }= \frac{L-N}{N+1} = \frac{1}{n^a} - 1 + \mathcal{O}(\nicefrac{1}{N}).
\end{equation}

It is worth to underline that the amplitude of the hopping in this new basis does not carry the usual Bose-enhancement factors, however, it requires extra factors $\hat{\rho}^{-1/2}_n$ in the Hamiltonian. Since the latter only show up in combination with $\hat{\Psi}_n$, the expression vanishes and remains well-defined when the bosonic occupation numbers become zero.

\emph{Field theory analysis.--}
Now we analyze the model \eqref{eqHtransfrm} from a field-theoretic perspective. By construction these models are connected by a unitary transformation (the non-local basis change), ensuring their spectra to coincide. At long wavelengths, distances are related as follows: $x$ in the \Zt LGT corresponds to a 'distance' (particle number) in the string-length basis $\tilde{x}=n^a x$. As a result we can directly relate coarse-grained densities in the two models.

This allows us to directly relate their Luttinger parameters $\tilde{K}$ and $K$, which can be defined via the compressibility \cite{Giamarchi2004}. An explicit calculation \cite{SMPhysRev} yields:
\begin{equation}
 K = (n^a)^2 ~ \tilde{K},
 \label{eqRelLuttingerK}
\end{equation}
reminiscent of the Luther-Emery re-scaling solution \cite{Giamarchi2004,Luther1974}, except for a factor of two.

Alternatively, we can relate density-density correlations at long-distances in the two models: we start from $\langle \delta \hat{n}(x) \delta \hat{n}(0) \rangle$, where $ \delta \hat{n}(x) = \hat{n}(x) - \left \langle \hat{n}(x) \right \rangle$ denotes local density-fluctuations. At long wavelengths, the density of hard-core bosons is $\hat{n}(x) \approx \Delta \hat{N}_a / \Delta x$, when $\Delta \hat{N}_a$ particles are found per coarse-grained distance $\Delta x$. In the string-length basis, $\hat{\rho}(\tilde{x})$ describes the distance $\Delta \hat{x}$ between two hard-core bosons, minus one unit per particle (Eq.~\eqref{eqDefStringBasis}), per coarse-grained number of particles $\Delta \tilde{x}$; i.e. $ \hat{\rho}(\tilde{x}) \approx  ( \Delta \hat{x} - \Delta \tilde{x}) / \Delta \tilde{x}$. This leads to 
\begin{equation}
\hat{n}(x) \approx \bigl[ 1 +  \hat{\rho}(\tilde{x}(x)) \bigr]^{-1},
\end{equation}
which allows us to calculate density fluctuations at long distances, $\delta \hat{n}(x) =  - (n^a)^2 \delta \hat{\rho}(\tilde{x}) + \mathcal{O}(\delta \hat{\rho}^2)$. Hence both models share the same long-wavelength correlations:
\begin{equation}
\langle \delta \hat{n}(x) \delta \hat{n}(0) \rangle \simeq (n^a)^4 \langle \delta \hat{\rho}(n^a x) \delta \hat{\rho}(0) \rangle.
\label{eqNNcorrRel}
\end{equation}

For the local Hamiltonian \eqref{eqHtransfrm} we can safely apply Luttinger-liquid theory, which yields \cite{Giamarchi2004}
\begin{equation}
\langle \delta \hat{\rho}(\tilde{x}) \delta \hat{\rho}(0) \rangle \simeq \frac{\tilde{K}}{2 \pi^2} \frac{1}{\tilde{x}^2} + \frac{(\rho^\Psi)^2}{2} \l \frac{\tilde{\alpha}}{\tilde{x}} \r^{2 \tilde{K}} \cos(2 \pi \rho^\Psi \tilde{x}) + ...
\label{DnstyCorrStrngLngth}
\end{equation}
where $\tilde{\alpha}$ is a non-universal short-distance cut-off. From Eq.~\eqref{eqNNcorrRel} we thus predict in the original model:
\begin{multline}
\langle \delta \hat{n}(x) \delta \hat{n}(0) \rangle \simeq \frac{\tilde{K} (n^a)^2 }{2 \pi^2 x^2} + ... = \\
=  \frac{K}{2 \pi^2 x^2} + \frac{(n^a)^{2}}{2} \l \frac{\alpha}{x} \r^{2K} \cos(2 \pi n^a x) + ...
\label{DnstyCorr}
\end{multline}
This result confirms the relation between Luttinger parameters stated earlier, see Eq.~\eqref{eqRelLuttingerK}. 

Note however that the relation \eqref{eqNNcorrRel} does \emph{not} correctly predict the power-law of the oscillatory part in the correlations, which involves large wavevectors $2 \tilde{k}_F = 2 \pi \rho^\Psi$. We believe this is directly related to the failure of naive bosonization arguments in predicting the correct long-wavelength behavior of the Green's function. Since $\cos (2 \pi \rho^\Psi \tilde{x}) = \cos (2 \pi (x - n^a x)) \equiv \cos (2 \pi n^a x)$, the period of the oscillations is correctly captured however. 
As shown in \cite{SMPhysRev} our field-theoretic arguments are confirmed by the behavior of the density-density correlations which, for $h=V=0$, we calculate by Monte-Carlo sampling of the resulting free fermion theory \cite{Prosko2017} in the string-length representation and  by DMRG calculations for finite $h$ and $V$. The resulting fits confirm the universal Luttinger liquid behaviors \eqref{DnstyCorrStrngLngth}, \eqref{DnstyCorr} and the predicted relation between the Luttinger parameters.

\emph{Confinement as translational symmetry breaking.--}
In the string-length basis, the gauge invariant Green's function $g^{(1)}(x)$ translates to a highly non-local operator. Its most important effect is to shift string-length labels $\ell_m \to \ell_{m+1}$ for particle numbers $m$ between $ \tilde{x}_1 < m < \tilde{x}_2$, where $\tilde{x}_2 - \tilde{x}_1 = \tilde{x} = n^a x$, i.e.:
\begin{equation}
 g^{(1)}(x) \simeq  \langle \hat{T}(0,\tilde{x})  \rangle, 
 \label{eqConfEstimate}
\end{equation}
where we define the partial translation operator:
\begin{multline}
 \hat{T}(\tilde{x}_1,\tilde{x}_2)  \ket{ ... \ell_{\tilde{x}_1-1} ~\ell_{\tilde{x}_1} ~ ~ ... ~ ~ \ell_{\tilde{x}_2-1}~ \ell_{\tilde{x}_2} ~\ell_{\tilde{x}_2+1} ... } =  \\
 =  \ket{ ... \ell_{\tilde{x}_1-1}~ \ell_{\tilde{x}_2} ~ \ell_{\tilde{x}_1} ~~...  ~~\ell_{\tilde{x}_2-1}~ \ell_{\tilde{x}_2+1} ... } 
\end{multline}
which cyclically shifts all string-occupations by one unit between 'sites' $\tilde{x}_1$ and $\tilde{x}_2$. 

Aside from local terms around $\tilde{x}_1$ and $\tilde{x}_2$, which can be assumed to yield non-zero additional factors and were thus neglected in Eq.~\eqref{eqConfEstimate}, the $g^{(1)}(x)$ function essentially probes translational invariance of the eigenstates in the string-length basis. Whenever the lattice translation symmetry $\tilde{x} \to \tilde{x}+1$ is broken throughout the system (spontaneously, or as in Eq.~\eqref{eqHtransfrm} by a non-zero field $h \neq 0$), it follows that
\begin{equation}
g^{(1)}(x) \simeq \langle \hat{T}(0,\tilde{x})  \rangle \simeq e^{ - \tilde{\kappa} \tilde{x} } = e^{ - \tilde{\kappa} n^a x },
\end{equation}
i.e. the corresponding \Zt LGT is confining. 

Using this argument, it is now easy to see that the original model in Eq.~\eqref{eqDefModel} \emph{must be confining for any $h \neq 0$.} Random $h_{\ij}$ would similarly lead to confinement. Even for $h=0$ it can become confining if translational symmetry is spontaneously broken by additional interactions: this case corresponds to a Mott insulating phase.

\emph{Mott transition and HODQLRO.--}
Earlier studies of the model \eqref{eqDefModel} have revealed no Mott insulating states in the absence of the repulsive NN interaction, $V=0$ \cite{Borla2020}. There, the model maps to free fermions for $h = 0$ \cite{Prosko2017} and the field $h \neq 0$ inducing confinement is not sufficient to reach the insulating state. In the limit where $h \rightarrow \infty$ and $V = 0$ the particles are bound in dimers and the effective model maps exactly to a 1D Heisenberg antiferromagnet. Further analysis showed that at a special filling $n^{a} = 2/3$ the system is at the critical point with $K=8/9$, right at the transition from the Luttinger liquid to a Mott insulating phase \cite{Borla2020}. In this regime the string-length model features HODQLRO since $\tilde{K} = 2 > 1$ \cite{SMPhysRev}.

In the following we will focus on the filling $n^{a} =2/3$, which corresponds to half-filing in the string-length basis, $\rho^{\Psi} = 1/2$. We consider the repulsive interaction $V \geq 0$ and show that it can stabilize the Mott insulator. Two limits are analytically tractable: For $h \rightarrow \infty$, infinitesimal $V>0$ opens a Mott gap. This is a BKT transition, as can be understood from the aforementioned mapping of the effective dimer model to the $SU(2)$ invariant Heisenberg model following \cite{Borla2020}. On the other hand, for $h = 0$ even $V \rightarrow \infty$ is insufficient to obtain the gapped state. However, as we show by an explicit calculation in \cite{SMPhysRev}, an infinitesimal $h \neq 0$ is sufficient to obtain a gapped phase when $V \rightarrow \infty$.

\begin{figure}[t!]
\centering
\epsfig{file=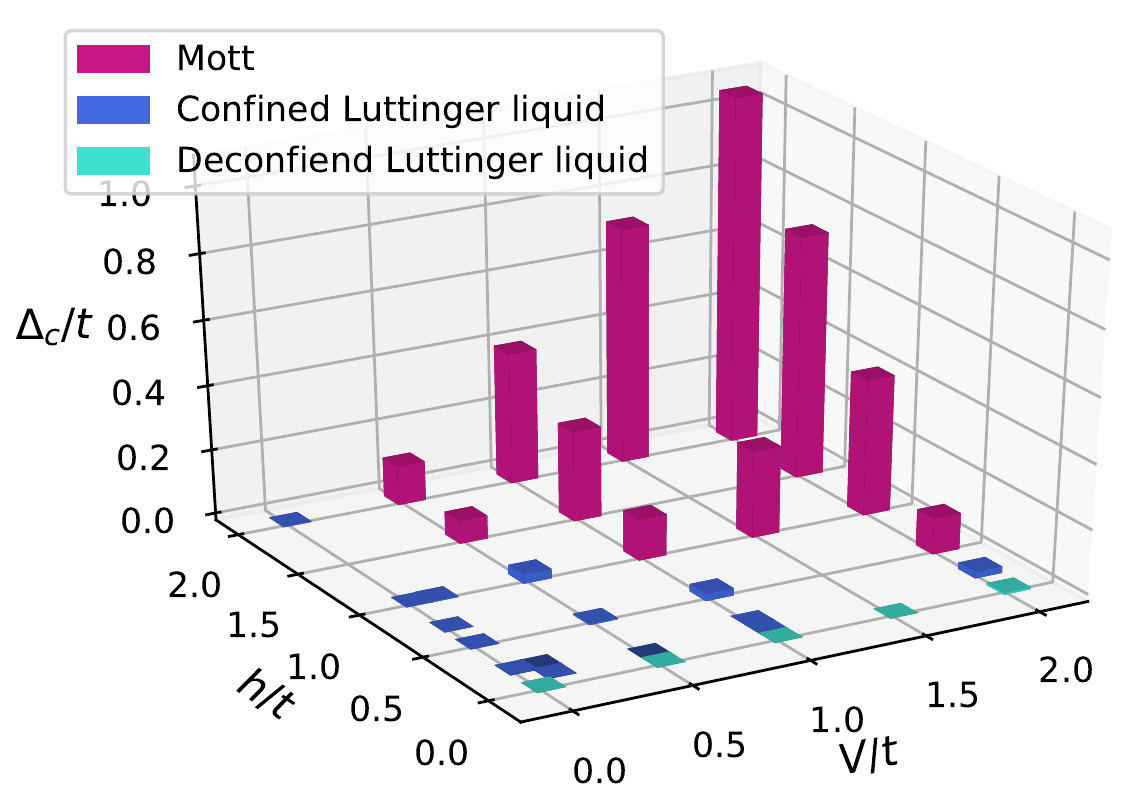, width=0.475\textwidth}
\caption{Charge gap extrapolated in the thermodynamic limit, $\Delta_{c}$, at filling $n^a=2/3$ as a function of $h$ and $V$. As a guide for the eye the violet (blue) bars denote $\Delta_{c}/t > 0.05$ ($\Delta_{c}/t \leq 0.05$) where we expect a Mott insulator (Luttinger liquid).}
\label{figGapBars}
\end{figure}

For generic nonzero values $h, V > 0$ we performed DMRG \cite{Schollwoeck2011, hubig:_syten_toolk, hubig17:_symmet_protec_tensor_networ} calculations to extract the charge gap
\begin{equation}
	\Delta_{c}(L, N) = \frac{1}{2} \bigl[ (E_{N+2}^{L} -E_{N}^{L}) - (E_{N}^{L} -E_{N-2}^{L}) \bigr],
	\label{eqGap}
\end{equation}
where $E_{N}^{L}$ is the ground state energy of the original \Zt LGT model with chain length $L$ and boson number $N$. We fixed the ratio of $N$ and $L$ at $n^{a} = \frac{N}{L} = 2/3$ and extrapolated the gap $\Delta_{c}$ in the thermodynamic limit by considering $L \rightarrow \infty $, see \cite{SMPhysRev}. As can be seen in Fig.~\ref{figGapBars} the Mott insulating state is reached only in the case when both parameters are nonzero $h, V \neq 0$ and large enough. For a fixed value of $V$ we observe an exponential opening of the gap as a function of $h$. The precise value of the transition point $h_c$ is difficult to extract, but the exponential behavior of the gap opening points to a BKT nature of the transition, see \cite{SMPhysRev}.

\emph{Discussion and outlook.--}
We have solved the confinement problem of dynamical charges in a class of 1D \Zt LGTs by means of a non-local string-length representation, which has revealed an unexpected relation to translational symmetry breaking. Our arguments should apply equally for other gauge groups in 1D. We found that, while the gauge symmetry keeps the Luttinger-liquid paradigm valid, the non-local interactions mediated by the gauge field must be treated with care. In particular, the confined gapless phase is characterized by an exponentially decaying \Zt invariant Green's function but we found that it features HODQLRO in the string-length basis before a confined Mott state is realized. 

We have analyzed the Mott insulating state at filling $n^{a} = 2/3$ and showed that it is stabilized by a combination of $h,V \neq 0$. An interesting future extension would be to consider the filling $n^{a} = 1/2$, where repulsive NN interactions $V \neq 0$ can readily stabilize a Mott insulator when $h=0$. On the other hand, one still finds a gapless system for $h \neq 0$ \cite{Borla2020} and a large $|h| \gg t$ is expected to destabilize the Mott insulator. Other extensions of our work include generalization to spin-full systems, higher dimensions and more complicated gauge groups.

\emph{Acknowledgements.--}
We thank U. Borla, S. Moroz, R. Verresen, N. Goldman, C. Schweizer, M. Aidelsburger, L. Pollet, F. Horn, F. Palm and S. Mardazad for fruitful discussions. This research was funded by the Deutsche Forschungsgemeinschaft (DFG, German Research Foundation) via Research Unit FOR 2414 under project number 277974659, and by the Deutsche Forschungsgemeinschaft (DFG, German Research Foundation) under Germany's Excellence Strategy -- EXC-2111 -- 390814868. M.~K. acknowledges the Ad Futura Scholarship (244. javni razpis) from the Public Scholarship, Development, Disability and Maintenance Found of the Republic of Slovenia.  L.~B. acknowledges support from Agencia Estatal de Investigación (“Severo Ochoa” Center of Excellence CEX2019-000910-S, Plan National FIDEUA PID2019-106901GB-I00/10.13039 / 501100011033, FPI), Fundació Privada Cellex, Fundació Mir-Puig, and from Generalitat de Catalunya (AGAUR Grant No. 2017 SGR 1341, QuantumCAT U16-011424, CERCA program) and from Topocold ERC starting grant.

%

\newpage

\section{Supplementary Material}

\subsection{Realization of the \Zt LGT by Rydberg dressing}
In the following we will show that the Hamiltonian obtained from the Rydberg dressing scheme \eqref{eqRydbergEffectiveNN} maps to the 1D LGT model \eqref{eqDefModel}, provided that we restrict our Hilbert space to a specific spin configuration.

Rydberg states can be used in cold atom experiments since they provide a way to precisely tune interaction between particles \cite{Saffman2010}. Due to high strength of such interactions it is sufficient to use the dressing scheme where the effective Ising interactions are proportional to $\frac{1}{r^6+a_{B}^6}$, where $a_{B}$ is the blockade radius \cite{Henkel2010}. Due to the aforementioned proportionality we can consider a situation where the on-site interaction is sufficiently large that doublon formation is forbidden. This requires $a_{b} < a$ where $a$ is the lattice spacing. On the other hand the interaction beyond NN can be neglected due to rapid decay of the potential, $V \propto 1/r^6$. 

\begin{figure}[t!]
\centering
\epsfig{file=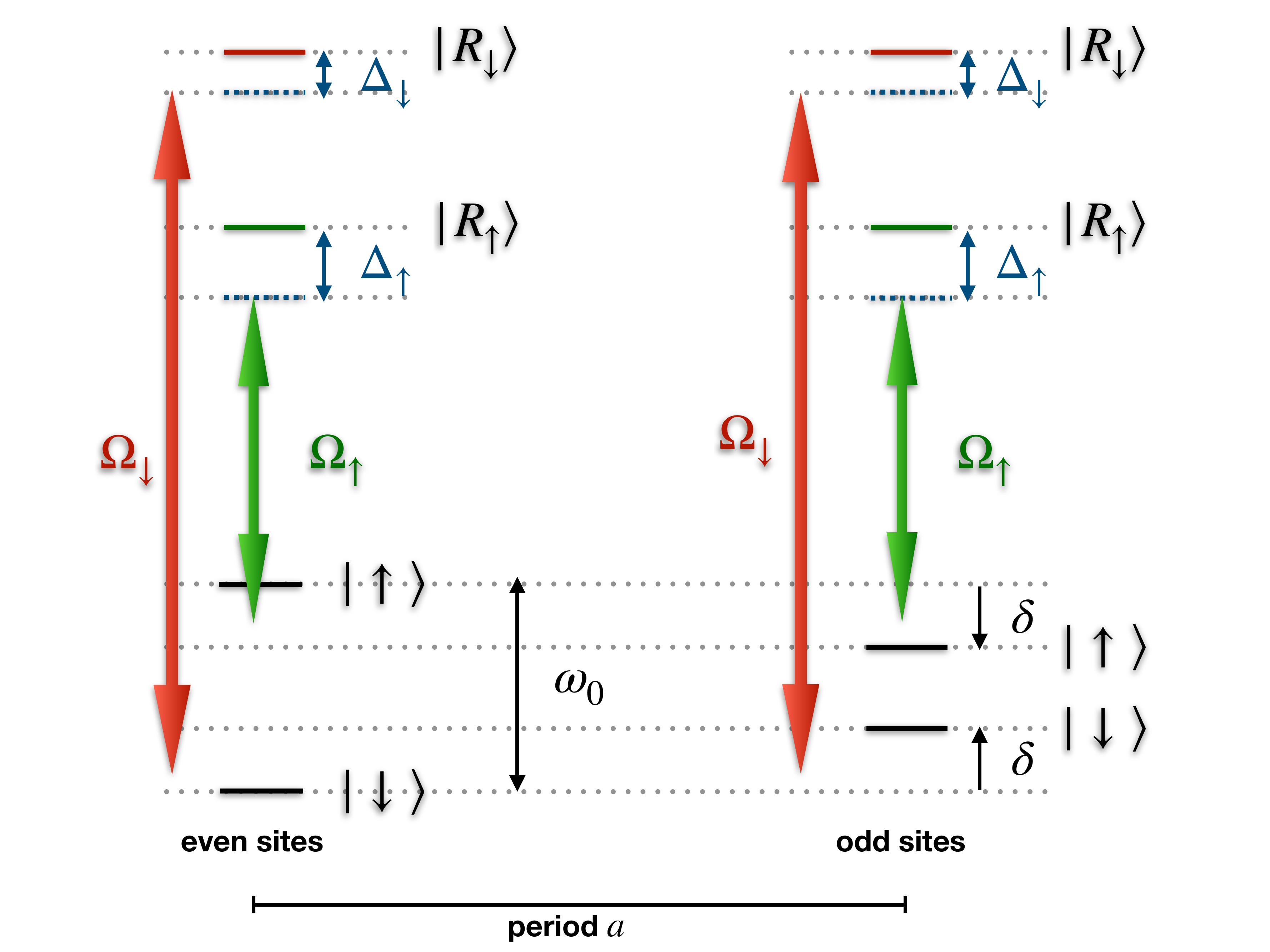, width=0.48\textwidth}
\caption{Proposed Rydberg dressing scheme. By using the anti-magic wavelength property of, e.g., Yb atoms one can achieve an effective potential Eq.~\eqref{eqRydbergEffectiveNN}. By using two different Rydberg dressing lasers $\Omega_{\uparrow}$ and $\Omega_{\downarrow}$, with a large detuning $\Delta_{\uparrow}, \Delta_{\downarrow} \gg \Omega_{\uparrow}, \Omega_{\downarrow}$ this setup ultimately maps to the NN interaction of Eq.~\eqref{eqDefModel}.} 
\label{figRydbergScheme}
\end{figure}

In order to implement the potential Eq.~\eqref{eqPotential} two Rydberg dressing lasers and an anti-magic potential for the atoms is needed (see Fig.~\ref{figRydbergScheme}) as discussed in the main text. 

The effective $t-J_z$ type Hamiltonian obtained in the Rydberg dressing scheme equals to
\begin{multline}
	\H_{t-J_{z}} = \H _{t} + \sum_{\langle i, j \rangle} \bigg ( J_{\uparrow \uparrow} \ket{\uparrow _{i} \uparrow _{j}} \bra{\uparrow _{i} \uparrow _{j}} +
	\\ + J_{\uparrow \downarrow} \big ( \ket{\uparrow _{i} \downarrow _{j}} \bra{\uparrow _{i} \downarrow _{j}} +  \ket{\downarrow _{i} \uparrow _{j}} \bra{\downarrow _{i} \uparrow _{j}} \big ) +
	\\ + J_{\downarrow \downarrow} \ket{\downarrow _{i} \downarrow _{j}} \bra{\downarrow _{i} \downarrow _{j}} \bigg ) + \delta \sum_{j} (-1)^{j} \hat{S}^{z}_{j}.
\label{eqRydbergEffective}
\end{multline}
We again use the projector, $\hat{P}$ which projects onto a Hilbert space with zero or single occupancy on each lattice site and explicitly rewrite the Hamiltonian
\begin{multline}
	\H_{t-J_{z}} =  -t  \sum_{\langle i, j \rangle , \sigma} \hat{P} \left( \hat{c}^{\dagger}_{i, \sigma} \hat{c}_{j, \sigma} + h.c. \right ) \hat{P} + \sum_{\langle i, j \rangle} \bigg ( J_{\uparrow \uparrow} \hat{n}_{i}^{\uparrow} \hat{n}_{j}^{\uparrow} +
	\\+ J_{\uparrow \downarrow} \big ( \hat{n}_{i}^{\uparrow} \hat{n}_{j}^{\downarrow} + \hat{n}_{i}^{\downarrow} \hat{n}_{j}^{\uparrow}  \big ) +  J_{\downarrow \downarrow} \hat{n}_{i}^{\downarrow} \hat{n}_{j}^{\downarrow}  \bigg ) + \delta \sum_{j} (-1)^{j} \hat{S}^{z}_{j},
\label{eqRydbergEffectiveProjected}
\end{multline}
where $\hat{c}^{\dagger}_{i, \sigma}$ is the fermion creation operator on site $i$ with spin $\sigma$ and $\hat{n}_{i, \sigma} = \hat{c}^{\dagger}_{i, \sigma} \hat{c}_{i, \sigma}$ is the fermion density operator with spin $\sigma$. Full density on site $i$ is hence $ \hat{n}_{i} = \sum_{\sigma \in \{ \uparrow, \downarrow \}} \hat{n}_{i, \sigma}$. This Hamiltonian conserves the spin pattern since it only contains a classical Ising interaction.

We now consider the so called squeezed space \cite{Ogata1990, Hilker2017} picture of our system. Squeezed space is defined as a system from which we remove all empty sites (holes) and accordingly relabel the lattice indices, while maintaining the initial order of the spin configuration. We constrain our system in the squeezed space to Néel states. As a result the spin configuration for zero doping is $\hat{S}^{z}_{j} = \frac{1}{2}(-1)^{j}$. For non-zero doping (in real space) each hole shifts the spins by one site, hence
\begin{equation}
	\hat{S}_{j}^{z} = \frac{1}{2}(-1)^{j} \big [ \prod_{i \leq j} (-1)^{\hat{n}_{i}^{h}} \big ] (1- \hat{n}_{j}^{h} ).
\label{eqSpinZtoLGT}
\end{equation}

This motivates us to define the non-local $\mathbb{Z}_{2}$ electric field $\hat{\tau}_{\langle j,j+1 \rangle}^{x}$ as
\begin{equation}
	\hat{\tau}_{\langle j,j+1 \rangle}^{x} = \prod_{i \leq j} (-1)^{\hat{n}_{i}^{h}},
\label{eqDefTauX}
\end{equation}
where the $\hat{n}_{i}^{h} = 1 - \hat{n}_{i}$ is the hole density operator. By definition $\hat{\tau}_{\langle j,j+1 \rangle}^{x}$ changes sign across each hole. Hence we obtain the Gauss law, as can be seen also from direct calculation
\begin{equation}\label{eqGauss}
\hat{G}_{i} = \hat{\tau}_{\langle i-1, i \rangle}^{x} \hat{\tau}_{\langle i, i+1 \rangle}^{x} (-1)^{\hat{n}_{i}^{h}} = (-1)^{\hat{n}_{i}^{h}}  (-1)^{\hat{n}_{i}^{h}} = 1
\end{equation}
and the obtained gauge sectors are $G_{j} = 1, \forall j$.

Now we will use relations \eqref{eqSpinZtoLGT} and \eqref{eqDefTauX} to show that the Hamiltonian \eqref{eqRydbergEffectiveProjected} maps to \eqref{eqDefModel}. We will do this term by term.
The staggered term is the easiest to show as we just have to make use of Eq.~\eqref{eqSpinZtoLGT} and substitute the product $\prod_{i \leq j} (-1)^{\hat{n}_{i}^{h}}$ with Eq.~\eqref{eqDefTauX}
\begin{multline}
	\delta \sum_{j} (-1)^{j} \hat{S}^{z}_{j} = \delta \sum_{j} (-1)^{j}  \frac{1}{2}(-1)^{j} \big [ \prod_{i \leq j} (-1)^{\hat{n}_{i}^{h}} \big ] (1- \hat{n}_{j}^{h} ) 
	\\ = \frac{1}{2} \delta \sum_{j} ( \hat{\tau}_{\langle j,j+1 \rangle}^{x} - \hat{\tau}_{\langle j,j+1 \rangle}^{x}\hat{n}_{j}^{h}).
\label{eq:MappSztoTauX}
\end{multline} 
by rewriting the hole operator as $\hat{n}_{i}^{h} =\frac{1}{2} (1- (-1)^{\hat{n}_{i}^{h}}) $ and rewriting the Gauss law as 
\begin{equation}
	\hat{\tau}_{\langle i-1,i \rangle}^{x} = \hat{\tau}_{\langle i,i+1 \rangle}^{x}(-1)^{\hat{n}_{i}^{h}} 
\label{eqGaussMod}
\end{equation}
we obtain the mapping
\begin{equation}\label{eq:FieldTermLGT}
	\delta \sum_{j} (-1)^{j} \hat{S}^{z}_{j} \rightarrow  h \sum_{\langle i, j \rangle}  \hat{\tau}_{\langle i, j \rangle}^{x},
\end{equation}
where $\delta = h$.

For the hopping term we employ the slave particle formalism \cite{Auerbach1994} where we write the annihilation operator as
\begin{equation}\label{eqSlaveP}
	\hat{c}_{i, \sigma} = \hat{h}^{\dagger}_{i} \hat{f}_{i, \sigma},
\end{equation}
with the constraint
\begin{equation}
	\sum_{\sigma} \hat{f}^{\dagger}_{i, \sigma} \hat{f}_{i, \sigma}  +  \hat{h}^{\dagger}_{i} \hat{h}_{i} = 1,
\label{eqSlavePConstraint}
\end{equation}
which restricts the Hilbert space to zero or singly occupied lattice sites, meaning that we can drop the projectors $\hat{P}$. We rewrite the hopping term in terms of slave particles
\begin{equation}
	\hat{P} \hat{c}^{\dagger}_{i, \sigma} \hat{c}_{j, \sigma} \hat{P} =  \hat{h}_{i} \hat{h}^{\dagger}_{j} \hat{f}^{\dagger}_{i, \sigma} \hat{f}_{j, \sigma} = \H_{ij} \hat{\mathcal{Z}}_{ij, \sigma},
\label{eqHoppTerm}
\end{equation}
where we defined $\hat{\mathcal{Z}}_{ij, \sigma} = \hat{f}^{\dagger}_{i, \sigma} \hat{f}_{j, \sigma}$ and $\H_{ij} = \hat{h}_{i} \hat{h}^{\dagger}_{j}.$
We know that hopping of a fermion to the right is equivalent to hopping of a hole to the left. It is therefore useful to study the effect of the operator \eqref{eqHoppTerm} on the Fock state $\hat{\tau}_{\langle j, j+1 \rangle}^{x} \ket{\psi} = \prod_{i \leq j} (-1)^{\hat{n}_{i}^{h}} \ket{\psi}=\alpha \ket{\psi}$
\begin{equation}
	\hat{\mathcal{H}}_{j+1j} \hat{\mathcal{Z}}_{j+1j, \sigma} \hat{\tau}_{\langle j, j+1 \rangle}^{x} \ket{\psi} = + \alpha \hat{\mathcal{H}}_{j+1j} \hat{\mathcal{Z}}_{j+1j, \sigma} \ket{\psi}.
\label{eqAppliedHoppingSecond}
\end{equation}
\begin{equation}
	\hat{\tau}_{\langle j, j+1 \rangle}^{x} \hat{\mathcal{H}}_{j+1j} \hat{\mathcal{Z}}_{j+1j, \sigma} \ket{\psi} = - \alpha \hat{\mathcal{H}}_{j+1j} \hat{\mathcal{Z}}_{j+1j, \sigma} \ket{\psi},
\label{eqAppliedHoppingFirst}
\end{equation}
which follows from Eq.~\eqref{eqDefTauX}, meaning that a movement of charge (hole) by one site flips the gauge field between the old and the new site. This leads us to the mapping 
\begin{equation}
	\hat{\mathcal{Z}}_{j+1j, \sigma} \rightarrow \hat{\tau}_{\langle j+1, j \rangle}^{z},
\label{eq:ZmapsToGauge}
\end{equation}
and consequently
\begin{equation}
	-t \sum_{\langle i, j \rangle , \sigma} \hat{P} \left(\hat{c}^{\dagger}_{i, \sigma} \hat{c}_{j, \sigma} + h.c.\right ) \hat{P} \rightarrow -t \sum_{\langle i, j \rangle } (\hat{h}^{\dagger}_{i} \hat{\tau}_{\langle i, j \rangle}^{z} \hat{h}_{j} + h.c.).
\label{eqHoppingtJzToLGT}
\end{equation}

Finally we need to show the mapping of the NN interaction. We use one of the definitions of the spin operator $\hat{S}_{i}^{z} = \frac{1}{2} \sum_{\sigma} (-1)^{\sigma} \hat{c}_{i, \sigma}^{\dagger}\hat{c}_{i, \sigma}$ and rewrite the density operators as
\begin{equation} \label{eqOccupationComponent}
\begin{split}
	\hat{n}_{i, \uparrow} &=\frac{1}{2} \hat{n}_{i} +  \hat{S}_{i}^{z} ,  \\
	\hat{n}_{i, \downarrow} &=\frac{1}{2} \hat{n}_{i} -  \hat{S}_{i}^{z}.
\end{split}
\end{equation}
Interaction terms in \eqref{eqRydbergEffectiveProjected} can be rewritten as
\begin{equation}
\begin{split}
	\hat{n}_{i}^{\uparrow} \hat{n}_{j}^{\downarrow} + \hat{n}_{i}^{\downarrow} \hat{n}_{j}^{\uparrow} &= \frac{1}{2} \hat{n}_{i} \hat{n}_{j} - 2 \hat{S}_{i}^{z} \hat{S}_{j}^{z}, \\
	\hat{n}_{i}^{\uparrow} \hat{n}_{j}^{\uparrow}  &=  \frac{1}{4} \hat{n}_{i}\hat{n}_{j} + \frac{1}{2} \hat{S}_{i}^{z}  \hat{n}_{j} + \frac{1}{2}  \hat{n}_{i} \hat{S}_{j}^{z} + \hat{S}_{i}^{z} \hat{S}_{j}^{z}, \\
	\hat{n}_{i}^{\downarrow} \hat{n}_{j}^{\downarrow}  &= \frac{1}{4} \hat{n}_{i}\hat{n}_{j} - \frac{1}{2} \hat{S}_{i}^{z}  \hat{n}_{j} - \frac{1}{2}  \hat{n}_{i} \hat{S}_{j}^{z} + \hat{S}_{i}^{z} \hat{S}_{j}^{z}.
\end{split}
\label{eqInteractionRewriten}
\end{equation}
Note that since we are considering the 1D case, we write $j = i+1$. We once again use Eqs.~\eqref{eqSpinZtoLGT} and \eqref{eqDefTauX} to write the spin operator as  $\hat{S}_{j}^{z} = \frac{1}{2}(-1)^{j} \hat{\tau}_{\langle j, j+1 \rangle}^{x} (1- \hat{n}_{j}^{h} )$ and use this to map the terms in \eqref{eqInteractionRewriten}
\begin{multline}
	\hat{S}^{z}_{i} \hat{S}^{z}_{i+1}= -\frac{1}{4} \hat{\tau}_{\langle i, i+1 \rangle}^{x} \hat{\tau}_{\langle i+1, i+2 \rangle}^{x} ( 1 - \hat{n}_{i}^{h} - \hat{n}_{i+1}^{h}  + \hat{n}_{i}^{h} \hat{n}_{i+1}^{h} ) =
	\\ =  -\frac{1}{4} (1- \hat{n}_{i}^{h})(1- \hat{n}_{i+1}^{h}),
\label{eqMappSzSz}
\end{multline} 
\begin{multline}
	\hat{S}_{i}^{z} \hat{n}_{i+1} + \hat{n}_{i} \hat{S}_{i+1}^{z} =
	\\ = \frac{(-1)^{i}}{2} (1-\hat{n}^{h}_{i}) (1-\hat{n}^{h}_{i+1}) (\hat{\tau}_{\langle i, i+1 \rangle}^{x}  - \hat{\tau}_{\langle i+1, i+2 \rangle}^{x} )=
	\\ = -\frac{(-1)^{i}}{8} \big ( 1 + \hat{\tau}_{\langle i, i+1 \rangle}^{x} \hat{\tau}_{\langle i+1, i+2 \rangle}^{x} +  \hat{\tau}_{\langle i-1, i \rangle}^{x} \hat{\tau}_{\langle i, i+1 \rangle}^{x}+ \\
 	\\ + \hat{\tau}_{\langle i-1, i \rangle}^{x} \hat{\tau}_{\langle i+1, i+2 \rangle}^{x} \big ) (\hat{\tau}_{\langle i, i+1 \rangle}^{x} - \hat{\tau}_{\langle i+1, i+2 \rangle}^{x}) = 0,
\label{eqMappSzni}
\end{multline}
where in the last lines of Eqs.~\eqref{eqMappSzSz} and \eqref{eqMappSzni}  we used relation $\hat{n}_{i}^{h} = \frac{1}{2} (1- \hat{\tau}_{\langle i-1, i \rangle}^{x} \hat{\tau}_{\langle i, i+1 \rangle}^{x})$ which is directly obtained from the Gauss Law.
Using $\hat{n}_{i} \hat{n}_{i+1} = (1-\hat{n}^{h}_{i})(1-\hat{n}^{h}_{i+1}) $ the relations \eqref{eqInteractionRewriten} become
\begin{equation}
\begin{split}
	\hat{n}_{i}^{\uparrow} \hat{n}_{j}^{\downarrow} + \hat{n}_{i}^{\downarrow} \hat{n}_{j}^{\uparrow} &= (1- \hat{n}_{i}^{h})(1- \hat{n}_{i+1}^{h}), \\
	 \hat{n}_{i}^{\uparrow} \hat{n}_{j}^{\uparrow}  &=  0, \\
	 \hat{n}_{i}^{\downarrow} \hat{n}_{j}^{\downarrow}  &= 0.
\label{eqInteractionMapped}
\end{split}
\end{equation}
The interaction term therefore maps to
\begin{multline}
	\sum_{\langle i, j \rangle} \bigg (  J_{\uparrow \uparrow} \hat{n}_{i}^{\uparrow} \hat{n}_{j}^{\uparrow}  + J_{\uparrow \downarrow} \big ( \hat{n}_{i}^{\uparrow} \hat{n}_{j}^{\downarrow} + \hat{n}_{i}^{\downarrow} \hat{n}_{j}^{\uparrow}  \big ) +
	\\ +  J_{\downarrow \downarrow} \hat{n}_{i}^{\downarrow} \hat{n}_{j}^{\downarrow}\bigg ) \rightarrow  \sum_{\langle i, j \rangle}V (1- \hat{n}_{i}^{h})(1- \hat{n}_{j}^{h}),
\label{eqInteractionMapping}
\end{multline}
where $V = J_{\uparrow \downarrow}$. Extra terms $V$ and $V \hat{n}_{i}^{h} $ in Eq.~\eqref{eqInteractionMapping} amount to a constant energy offset and contribute to chemical potential.

\subsection{Relation of Luttinger $K$'s -- compressibilities}
The Luttinger $K$-parameter of a one-dimensional quantum liquid is related to its compressibility $\kappa$ by \cite{Giamarchi2004}
\begin{equation}
 \kappa \equiv - \frac{1}{L} \l \frac{\partial L}{\partial P} \r_{N} = \frac{1}{(n^a)^2} \frac{K}{\pi u^a}
 \label{eqKappaDefK}
\end{equation}
where $n^a$ is the corresponding density, $u^a$ denotes its speed of sound and $P = - \l \frac{\partial E}{\partial L} \r_{N}$ is the pressure with $E$ being the energy. Applying the general equation above to the string-length representation of the \Zt LGT yields for the corresponding compressibility
\begin{equation}
 \tilde{\kappa} = \frac{1}{( \rho^\Psi )^2} \frac{\tilde{K}}{\pi \tilde{u}},
 \label{eqTildeKappaDefK}
\end{equation}
where $\tilde{K}$, $\rho^\Psi$ and $\tilde{u}$ denote the respective Luttinger parameters, density and speed of sound in the string-length basis. 

We can relate the two Luttinger parameters $K$ and $\tilde{K}$ by using the geometric relations characterizing the mapping from the \Zt LGT to the string-length basis. We start by noticing that 
\begin{equation}
 \kappa = - \frac{1}{L} \l \frac{\partial L}{\partial P} \r_{N} = \left[ L \l \frac{\partial^2 E}{\partial L^2} \r_N \right]^{-1},
 \label{eqKappaDefPartial}
\end{equation}
and similarly 
\begin{equation}
 \tilde{\kappa} = \left[ \tilde{L} \l \frac{\partial^2 E}{\partial \tilde{L}^2} \r_{\tilde{N}} \right]^{-1},
 \label{eqTildeKappaDefPartial}
\end{equation}
where the energy $E$ coincides in both representations since they are related by a unitary transformation. In the following we will relate $\tilde{\kappa}$ to $\kappa$.

The 'particle number' $\tilde{N}$ and 'system size' $\tilde{L}$ in the string-length model are related to $N$ and $L$ in the \Zt LGT model by 
\begin{equation}
N(\tilde{N}, \tilde{L}) = \tilde{L}, \qquad L(\tilde{N}, \tilde{L}) = \tilde{N} + \tilde{L}.
\end{equation}
Hence for any function $A(\tilde{L}, \tilde{N})$
\begin{multline}
\l \frac{\partial A}{\partial \tilde{L}} \r_{\tilde{N}} = \l \frac{\partial A}{\partial L} \r_{N}  \l \frac{\partial L}{\partial \tilde{L}} \r_{\tilde{N}} + \l \frac{\partial A}{\partial N} \r_{L}  \l \frac{\partial N}{\partial \tilde{L}} \r_{\tilde{N}}\\
 = \l \frac{\partial A}{\partial L} \r_{N} + \l \frac{\partial A}{\partial N} \r_{L}.
\end{multline}
Applying this relation twice on the right hand side of \eqref{eqTildeKappaDefPartial} thus yields
\begin{equation}
 \l \frac{\partial^2 E}{\partial \tilde{L}^2} \r_{\tilde{N}} = \l \frac{\partial^2 E}{\partial L^2} \r_N + 2 \l \frac{\partial^2 E}{\partial L \partial N} \r + \l \frac{\partial^2 E}{\partial N^2} \r_L.
\end{equation}
Using the following Maxwell relations for the chemical potential $\mu =( \partial E / \partial N)_L$,
\begin{equation}
\l \frac{\partial \mu}{\partial N} \r_L  \! = - \frac{L}{N} \l \frac{\partial \mu}{\partial L} \r_N \! = \frac{L}{N} \l \frac{\partial P}{\partial N} \r_L \! = - \frac{L^2}{N^2} \l \frac{\partial P}{\partial L} \r_N,
\end{equation}
we can simplify
\begin{flalign}
\l \frac{\partial^2 E}{\partial L \partial N} \r &= \l \frac{\partial \mu}{\partial L} \r_N = \frac{L}{N} \l \frac{\partial P}{\partial L} \r_N, \\
\l \frac{\partial^2 E}{\partial N^2 } \r_L &= \l \frac{\partial \mu}{\partial N} \r_L = - \frac{L^2}{N^2} \l \frac{\partial P}{\partial L} \r_N, \\
\l \frac{\partial^2 E}{\partial L^2 } \r_N &= -  \l \frac{\partial P}{\partial L} \r_N.
\end{flalign}
Combining these results we obtain
\begin{equation}
 \l \frac{\partial^2 E}{\partial \tilde{L}^2} \r_{\tilde{N}}  = -  \l \frac{\partial P}{\partial L} \r_N \left[ 1 - 2 \frac{L}{N} + \frac{L^2}{N^2} \right].
 \label{eqd2Ed2Ltld}
\end{equation}

Next we plug Eq.~\eqref{eqd2Ed2Ltld} into Eq.~\eqref{eqTildeKappaDefPartial}. Using the definitions of the densities, $n^a = N/L$ and
\begin{equation}
\rho^\Psi = \frac{\tilde{N}}{\tilde{L}} =  \frac{1}{n^a} - 1,
\end{equation}
as well as Eq.~\eqref{eqKappaDefPartial} yields:
\begin{multline}
\tilde{\kappa} = - \left[ L \frac{\tilde{L}}{L}  \l \frac{\partial P}{\partial L} \r_N \l 1 - \frac{1}{n^a} \r^2 \right]^{-1} \\
 = - \frac{L}{\tilde{L}} (\rho^\Psi)^{-2} \left[  L  \l \frac{\partial P}{\partial L} \r_N \right]^{-1} \\
 = (\rho^\Psi)^{-2} (n^a)^{-1} \kappa.
\end{multline}
In the last step we used that $L / \tilde{L} = L/N = 1/ n^a$. Hence:
\begin{equation}
\frac{\tilde{\kappa}}{\kappa} = \frac{1}{n^a (\rho^\Psi)^2}.
\label{eqRatioKappa}
\end{equation}

We can also use the relations \eqref{eqKappaDefK}, \eqref{eqTildeKappaDefK} of the compressibilities to the Luttinger parameters to obtain
\begin{equation}
 \frac{\tilde{\kappa}}{\kappa} = \frac{\tilde{K}}{K} \l \frac{n^a}{\rho^\Psi} \r^2 \frac{u}{\tilde{u}}.
 \label{eqRatioKappaKs}
\end{equation}
Since 'distances' $\tilde{x}$ in the string-length representation are related to distances $x$ in the \Zt LGT by $\tilde{x} = n^a x$, the speed of sound $\tilde{u}$ is related to $u^a$ by
\begin{equation}
 \tilde{u} = \frac{d \tilde{x}}{d t} = n ^a \frac{d x}{dt} = n^a u.
 \label{eqTildeUvsU}
\end{equation}

Finally, combining Eqs.~\eqref{eqRatioKappa}, \eqref{eqRatioKappaKs} and \eqref{eqTildeUvsU} yields
\begin{equation}
\frac{\tilde{K}}{K} = \frac{1}{(n^a)^2},
\end{equation}
as claimed in the main part of the paper.

\subsection{Luttinger parameter for the confined phase}

In the limit where $h/t \rightarrow \infty$ and $V = 0$ the original model \eqref{eqDefModel} can be mapped to an effective dimer model, where confined pairs are squeezed into single particles \cite{Borla2020}. The number of particles in the effective dimer model is thus $N_{b} = N / 2$ and the length of the chain in the dimer model is $L_{b} =~L - N / 2$. Hence the density in the dimmer model can be expressed in terms of the density in the original model as $n^{b} = \frac{n^a}{2-n^a}$. Similarly the distance in the dimer model is $x^{b} = x - \frac{n^a}{2} x$. Using similar procedure as in the main text we obtain $\delta \hat{n}(x) = \frac{1}{2} \left( 2-n^{a} \right)^2 \delta \hat{n}^{b}(x^{b})$ and write the density-density correlations as
\begin{equation}
	\langle \delta \hat{n}(x) \delta \hat{n}(0) \rangle \simeq \frac{1}{4}(2-n^a)^4 \langle \delta \hat{n}^{b}(x^{b}) \delta \hat{n}^{b}(0) \rangle.
\end{equation}
Comparing the Luttinger parameters on both sides yields 
\begin{equation}
	K = \left(2-n^a \right)^2 K^{b},
	\label{eqLuttHeisenbergLGT}
\end{equation}
where $K^{b}$ is the Luttinger parameter in the effective dimer model.

The effective dimer model can be mapped to a \mbox{spin-1/2} Heisenberg model \cite{Borla2020}. Here the magnetization of the system is directly related to the filling of the original system and is zero for $\n^{a} = 2/3$. Hence the Luttinger parameter is equal to $K^{b} = 1/2$ \cite{Giamarchi2004}. Using the Eq.~\eqref{eqLuttHeisenbergLGT} the Luttinger parameter in our original model thus equals to $K = 8/9$ for $n^{a} = 2/3$.

\subsection{Density correlations: Luttinger liquid fits}

\emph{Monte-Carlo sampling of free fermions.--}
As a first step to verify our Luttinger liquid calculations we compared the Monte-Carlo sampled free-fermion and string-length basis density-density correlations. In particular we were interested whether the results in two different bases adhere to Eq.~\eqref{eqNNcorrRel}. To this end we calculated free fermion density-density correlations, $\langle \delta \hat{n}(x) \delta \hat{n}(0) \rangle$, and by using the mapping in Eq.~\eqref{eqDefStringBasis} the string-length density-density correlations, $\langle \delta \hat{\rho}(\tilde{x}) \delta \hat{\rho}(0) \rangle$.
\begin{figure}[h]
\centering
\epsfig{file=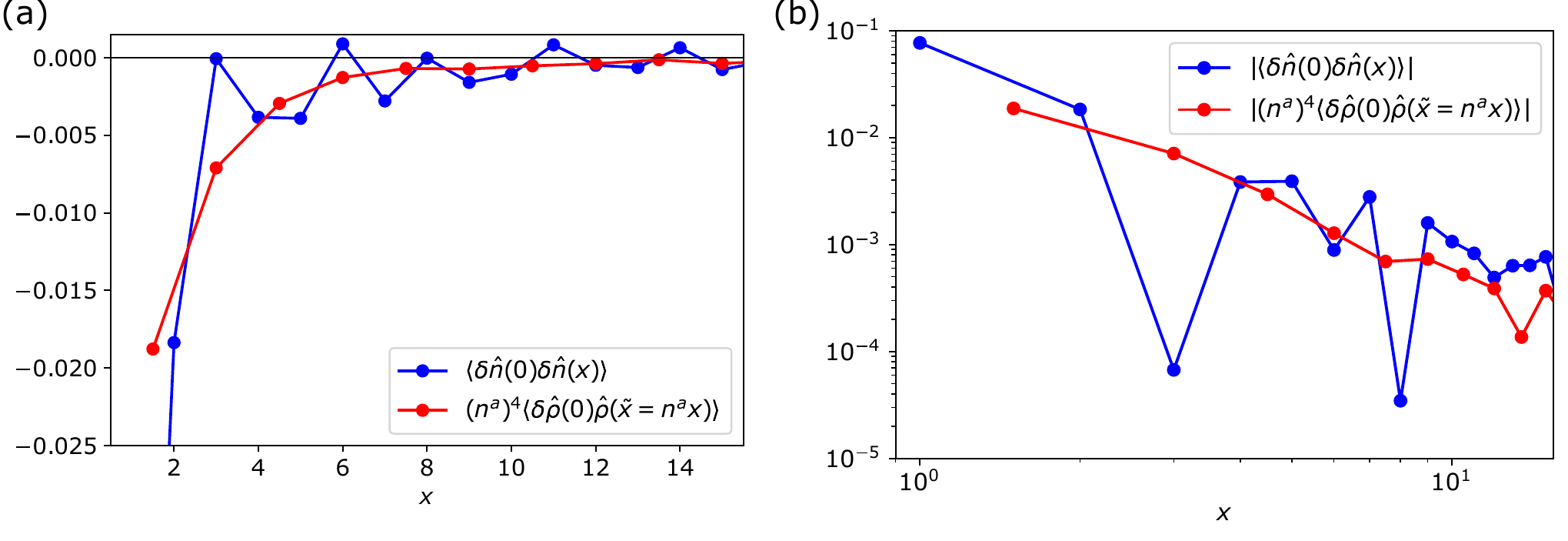, width=0.48\textwidth}
\caption{Density-Density correlations $\langle \delta \hat{n}(x) \delta \hat{n}(0) \rangle$ (blue data points) together with $ (n^{a})^{4} \langle \delta \hat{\rho}(x) \delta \hat{\rho}(0) \rangle$ (red data points) plotted in a linear scale (a) and in a log-log scale (b) as a function of $x$. We consider free fermions, $h=0$, at filling $n^a = 2/3$ and $\rho^{\Psi} = 1/2$.} 
\label{figMonteCarloComparison}
\end{figure}
All calculations were performed for $n^a = 2/3$, for which we simulated 240 fermions on 360 lattice sites.

The results of our calculations can be seen in Fig.~\ref{figMonteCarloComparison}. String-length data points were multiplied by $ (n^{a})^{4}$ and the $\tilde{x}$ had to be rescaled as $ \tilde{x}  \rightarrow x = \tilde{x}/n^{a} $ for a direct comparison. The data points for larger values acquired substantial relative error and hence only the first 15 data points are shown. High oscillations in the free-fermion basis, make the comparison a bit difficult, however a good overall agreement between the two bases can be seen.

\begin{figure}[t!]
\centering
\epsfig{file=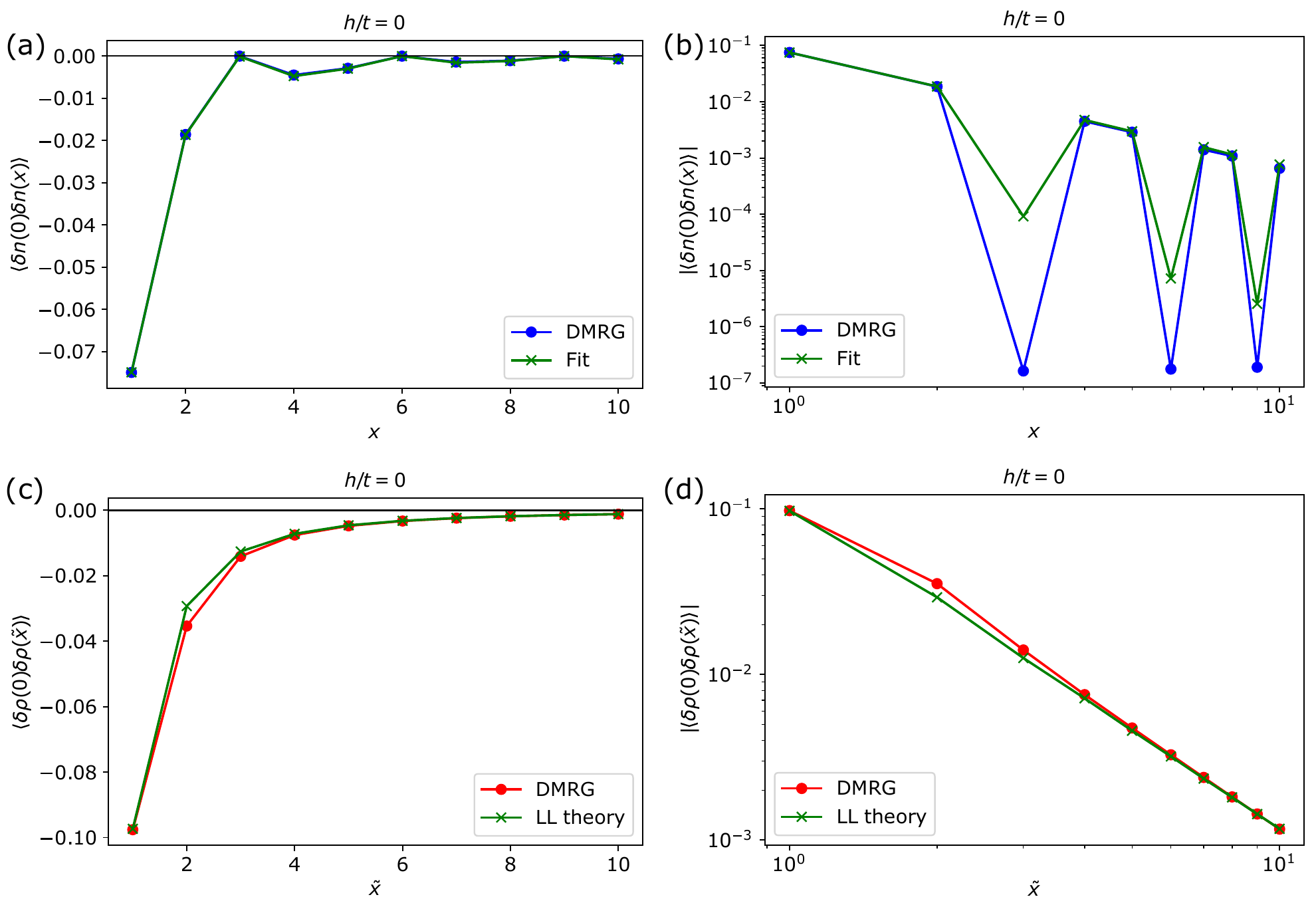, width=0.48\textwidth}
\caption{Density-density correlation in the original basis (blue dots) as a function of distance $x$ and the corresponding fitting function, $f(x)= A -\frac{K}{2 \pi^2 x^2} + \frac{(n^a)^{2}}{2} \l \frac{\alpha}{x} \r^{2K} \cos(2 \pi n^a x)$ (green data points) in linear (a) and log-log scaled axes (b). Density-density correlation in the string-length basis (red dots) as a function of distance $\tilde{x}$ and the corresponding fitting function, $f(\tilde{x})_{s}= A_{s} -\frac{\tilde{K}}{2 \pi^2 \tilde{x}^{2}} - \frac{(\rho^\Psi)^{2}}{2} \l \frac{\alpha}{\tilde{x}} \r^{2 \tilde{K}} \cos(2 \pi \rho^{\Psi} \tilde{x})$ (green data points), where we fixed the Luttinger parameter, $\tilde{K} = 2.25 $ in linear (c) and log-log scaled axes (d). We consider free fermions, $h=0$, at filling $n^a = 2/3$ and $\rho^{\Psi} = 1/2$.} 
\label{figOriginalChargeChargeh0}
\end{figure}

\emph{DMRG density-density correlations.--} For zero field regime, $h=0$, the original 1D \Zt LGT model maps to a free fermion model \cite{Prosko2017}. We thus expect the Luttinger parameter $K$ to be equal to unity. Using the DMRG (see section D) we calculated the charge-charge correlations for chain length $L = 210$ and fitted the results using Eq.~\eqref{DnstyCorr} with an added constant offset, see Fig.~\ref{figOriginalChargeChargeh0} (a). The resulting fit yields $K = 0.994 \pm 0.006, \alpha = 0.467 \pm 0.002, n = 0.6661 \pm 0.0008$ and $A = 5 \cdot 10^{-6} \pm 5.4 \cdot 10^{-5}$, where $A$ is a constant offset added to Eq.~\eqref{DnstyCorr} and the errors are square root values of the diagonal values of the covariance matrix of the fit. The parameters $n$ and $A$ were restricted to values close to $2 / 3$ and $ -10^{-5} < A < 5 \cdot 10^{-6}$ respectively. As can be seen in Fig.~\ref{figOriginalChargeChargeh0} (b) the amplitude of oscillations is not perfectly captured, most probably due to high sensitivity of Eq.~\eqref{DnstyCorr} to the value of $\alpha$.

\begin{figure}[t!]
\centering
\epsfig{file=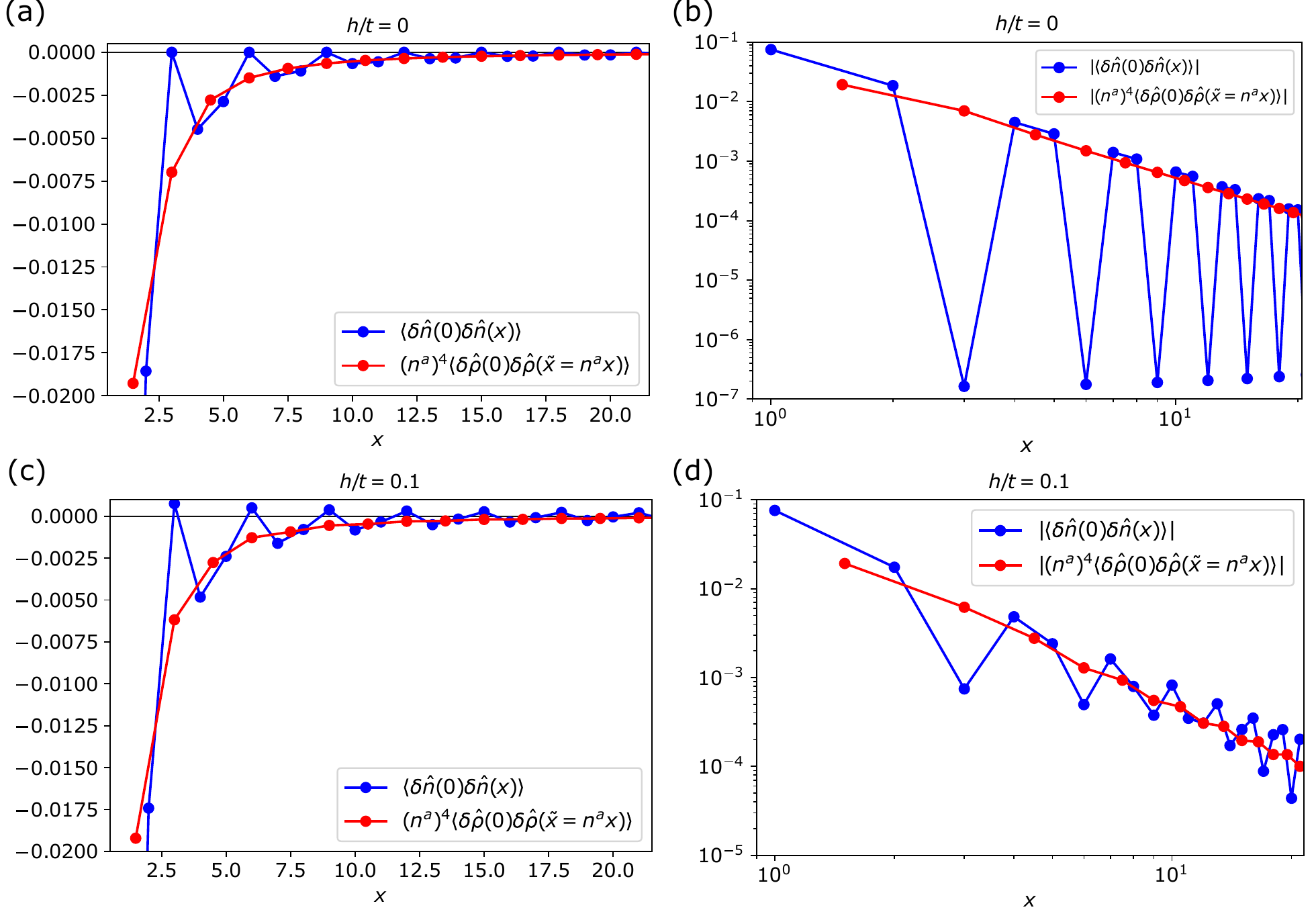, width=0.48\textwidth}
\caption{Comparison of DMRG simulations of the density-density correlations of the original \Zt LGT model (blue data points) and the rescaled charge-charge correlations, $ (n^{a})^{4} \langle \delta \hat{\rho}(x) \delta \hat{\rho}(0) \rangle$ in the string-length basis model (red data points). Aside from the oscillations for the original model a nice agreement can be seen for the zero field case, $h = 0$ in (a) and (b). Similar behaviour is found for nonzero field value, $h/t = 0.1$ which can be seen in (c) and (d). We consider filling $n^a = 2/3$ and $\rho^{\Psi} = 1/2$.}
\label{figDMRGComparison}
\end{figure}

We also performed DMRG simulations of the string-length Hamiltonian \eqref{eqHtransfrm}, and performed a similar fit for the density-density correlation function. Fixing the Luttinger-liquid parameter to the expected value $\tilde{K} = 2.25$ according to Eq.~\eqref{eqRelLuttingerK} and the density to $\rho^\Psi = 1/2$, we were able to fit the remaining parameters to our data with Eq.~\eqref{DnstyCorrStrngLngth}, again with an added constant offset, see Fig~\ref{figOriginalChargeChargeh0} (c). The resulting value for the fitting parameters were $\tilde{\alpha} = 0.64 \pm 0.11$ and $A_{s} = -3\cdot10^{-5} \pm 0.001$ where we again constrained the constant offset $-3\cdot10^{-5}<A_{s}<3\cdot10^{-5}$. Due to larger value of $\tilde{K}$ the oscillations are suppressed and are not as prominent as in the original basis. Nevertheless the slope of the fitting function agrees well with the data points presented in Fig~\ref{figOriginalChargeChargeh0} (d).

We also made a direct comparison of the data in the original and in the string-length basis for zero field, $h=0$ and for a nonzero field-value $h/t = 0.1$, see Fig.~\ref{figDMRGComparison}. Apart from large oscillations the slope of the original density-density correlator and the rescaled density-density correlator $(n^{a})^{4} \langle \delta \hat{\rho}(x) \delta \hat{\rho}(0) \rangle$,  $ \tilde{x}  \rightarrow x = \tilde{x}/n^{a} $ are similar which is in agreement with Eq.~\eqref{eqNNcorrRel}.

\begin{figure}[t!]
\centering
\epsfig{file=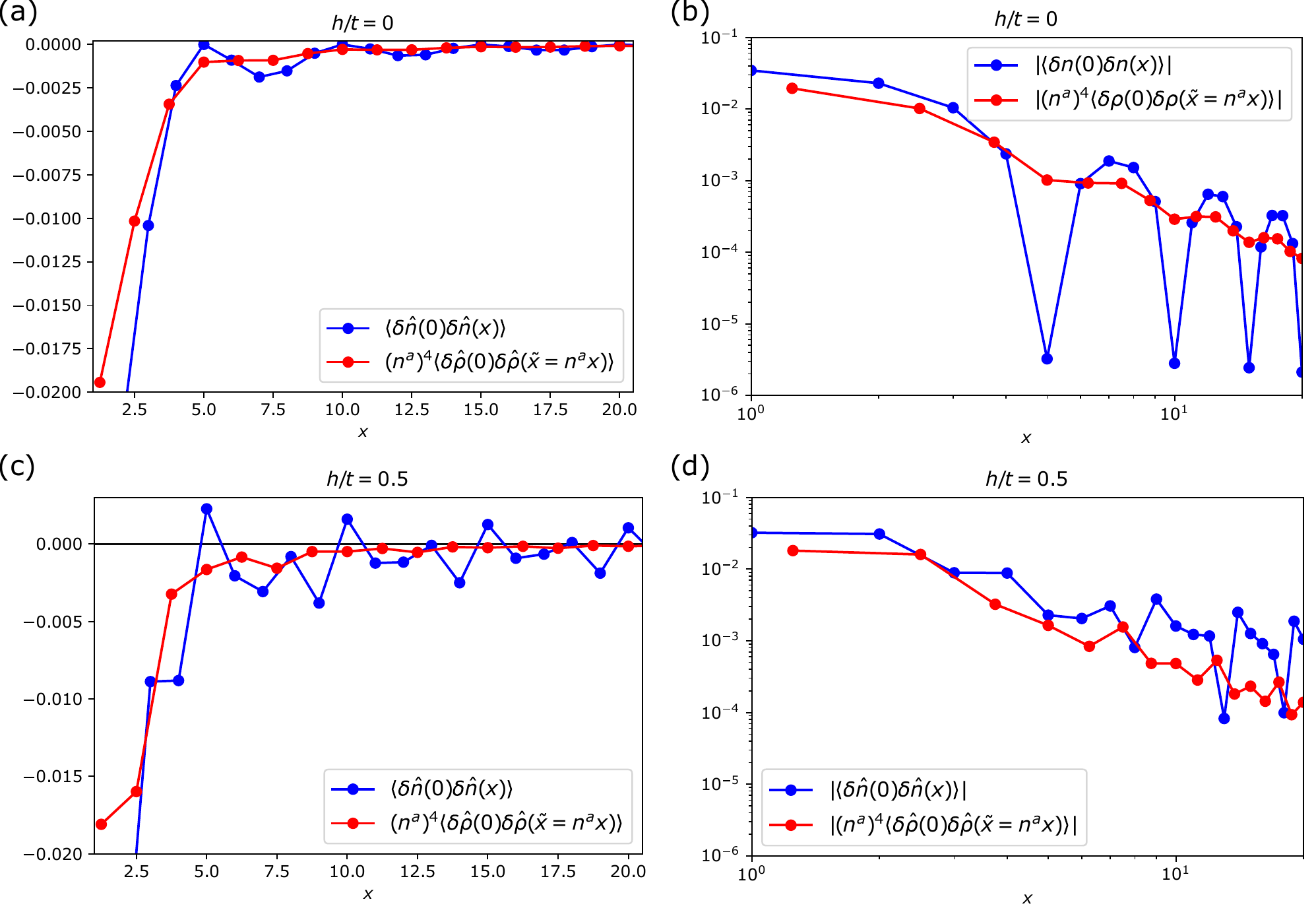, width=0.48\textwidth}
\caption{Comparison of DMRG simulations of the density-density correlations of the original \Zt LGT model $n^{a} = 4/5$ (blue data points) and the rescaled charge-charge correlations, $ (n^{a})^{4} \langle \delta \hat{\rho}(x) \delta \hat{\rho}(0) \rangle$ in the string-length basis model at $\rho^{\Psi} = 1/4$ (red data points). A nice agreement can be seen for the zero field case, $h = 0$ in (a) and (b) where in (a) the axes are scaled linearly and in (b) we plotted the same data in a log-log scale. Similar behaviour survives also for the first few data points for nonzero field value, $h/t = 0.5$ which can be seen in (c) for linear axes and in (d) for log-log scaled axes.}
\label{figDMRGComparisonDen08}
\end{figure}

In addition we also calculated the density-density correlations at filling of $n^a = 4/5$. We find good agreement for $h = 0$ and a good initial agreement for $h>0$, see Fig.~\ref{figDMRGComparisonDen08}. When we increase $h$ the first few data points coincide nicely, but for higher values of $x$ the correlations start to deviate, see Fig.~\ref{figDMRGComparisonDen08} (c) and (d).

This can be understood by closer examination of the scaling of the two correlation functions. We can see that the first terms of the correlation function Eq.~\eqref{DnstyCorrStrngLngth} and Eq.~\eqref{DnstyCorr} decay as  $\sim x^{-2} $ whereas the second term decays as $\sim x^{-2K} $ and $\sim x^{-2\tilde{K}} $ respectively. The latter, oscillatory terms, clearly deviate from the scaling \eqref{eqNNcorrRel}, as was already discussed in the main text. Moreover we see that the value of $K$ for the original model drops, $K<1$, as $h$ is increased. When $K<1$, the oscillatory part becomes dominant at long distances (quasi crystallization). This is noticeable in the linear plots but difficult to see in the log-log plot in Fig.~\ref{figDMRGComparisonDen08}. On the other hand the corresponding string-length exponent is considerably higher: E.g. for $n^{a} = 2/3$, the Luttinger parameter is $\tilde{K} \approx 2.25>1$ and the oscillations are barely visible, see Fig.~\ref{figDMRGComparison} (c) and (d). Hence we find quasi-condensation in the string-length basis. If we increase $n$, e.g., $n= 4/5$, the scaling $K = \left ( n^{a} \right )^{2} \tilde{K}$ ensures lower difference between both Luttinger parameters. Hence the oscillations in the string-length basis become visible already for $h=0$ (see Fig.~\ref{figDMRGComparisonDen08} (a) and (b) ) and become prominent as we increase $h$, since $\tilde{K}$ drops closer to unity, see Fig.~\ref{figDMRGComparisonDen08} (c) and (d).

\begin{figure}[t!]
\centering
\epsfig{file=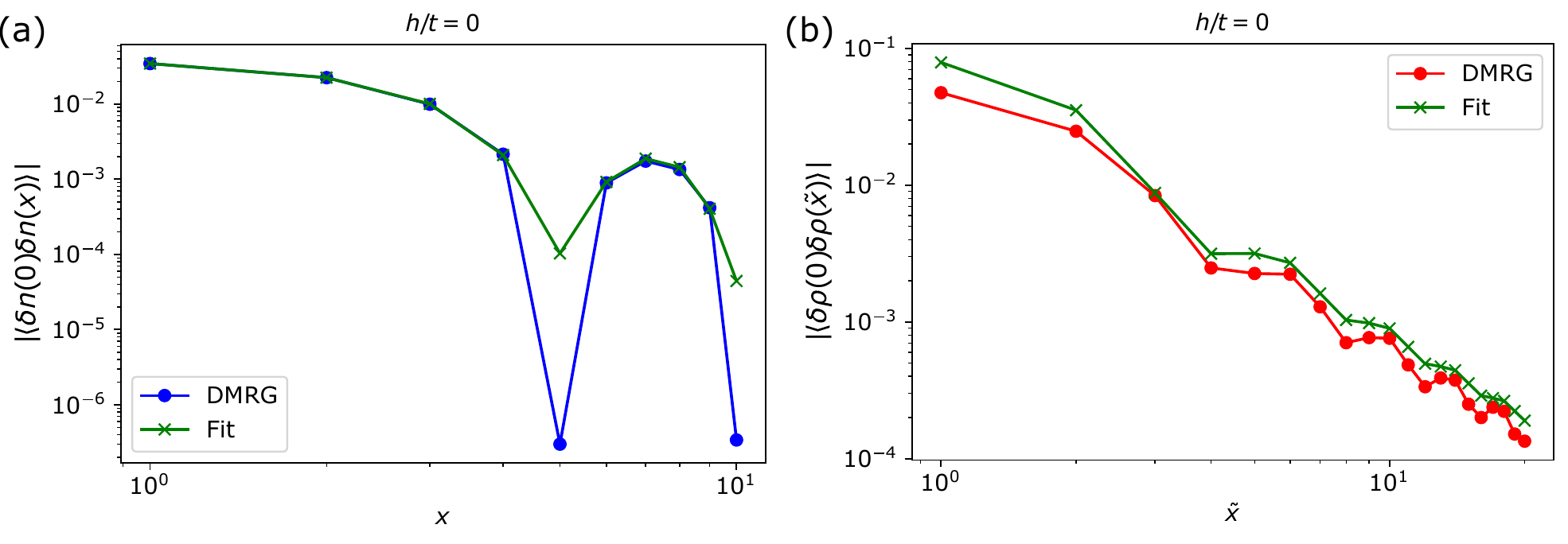, width=0.48\textwidth}
\caption{(a) Density-density correlation in the original basis, (blue dots) at filling $n^{a} = 4/5$, as a function of distance $x$ and the corresponding fitting function, $f(x)= A -\frac{K}{2 \pi^2 x^2} + \frac{(n^a)^{2}}{2} \l \frac{\alpha}{x} \r^{2K} \cos(2 \pi n^a x)$ (green data points). (b) Density-density correlation in the string-length basis, (red dots) at filling $\rho^{\Psi} = 1/4$ , as a function of distance $\tilde{x}$ and the corresponding function, $f(\tilde{x})_{s}= A_{s} -\frac{\tilde{K}}{2 \pi^2 \tilde{x}^{2}} + \frac{(\rho^\Psi)^{2}}{2} \l \frac{\alpha}{\tilde{x}} \r^{2 \tilde{K}} \cos(2 \pi \rho^{\Psi} \tilde{x})$ (green data points), where we fixed the parameters to $\tilde{K} = 1.56 $ , $A_{s} = -4 \cdot 10^{-6}$and $\tilde{\alpha} = 1.6$ (b). In both cases the respective densities were fixed paramters.} 
\label{figOriginalChargeChargeDen08h0}
\end{figure}

Precise fits are hard to obtain due to high sensitivity to $\alpha$ and $\tilde{\alpha}$. The fits at $h = 0$ for the original model, Fig.~\ref{figOriginalChargeChargeDen08h0} (a) yield $K = 0.976 \pm 0.008$ which is again close to the expected $K = 1$. The string-length basis was again difficult to fit. However, we compare the DMRG data with the curve obtained by using Eq.~\eqref{DnstyCorrStrngLngth} and inserting the expected Luttinger parameter, $\tilde{K} = 1.56$. Tuning the value of $\tilde{\alpha}$ and $A_{s}$ yields convincing agreement, see Fig.~\ref{figOriginalChargeChargeDen08h0}. 

\subsection{DMRG simulations}

\emph{Simulations of the \Zt LGT.--}
DMRG calculations of the \Zt LGT model were performed by using the \textsc{SyTen} toolkit created by Claudius Hubig \cite{Schollwoeck2011, hubig:_syten_toolk, hubig17:_symmet_protec_tensor_networ}.
The \Zt LGT model Eq.~\eqref{eqDefModel} was mapped to a spin-1/2 model by using the Gauss law, which was fixed to a sector where $G_{i} = 1$ on every lattice site. Using this result the $\mathbb{Z}_{2}$ electric field configuration on the links between lattice sites completely determines the hard-core boson configuration on the sites,
\begin{equation}
	\hat{n}_{j} = \frac{1}{2}(1- \hat{\tau}^x_{\ij} \hat{\tau}^x_{\langle j, k \rangle }).
	\label{eqNumberGaugeFields}
\end{equation}
By writing the Pauli matrices in terms of the spin-1/2 operators $\hat{\tau}^x_{\langle j, j+1 \rangle} \rightarrow 2 \hat{S}^{x}_{j}$ and $\hat{\tau}^z_{\langle j, j+1 \rangle} \rightarrow 2 \hat{S}^{z}_{j}$ the original \Zt LGT model becomes
\begin{multline}
	\hat{H}_{s} = t\sum_{i} (4\hat{S}^{x}_{i-1}\hat{S}^{x}_{i+1}-1) \hat{S}^{z}_{i} - 2h \sum_{i} \hat{S}^{x}_{i} \\
	+ V \sum_{i} \frac{1}{4} (1-4\hat{S}^{x}_{i-1}\hat{S}^{x}_{i}) (1-4\hat{S}^{x}_{i}\hat{S}^{x}_{i+1}) + \mu \sum_{i} \hat{S}^{x}_{i+1} \hat{S}^{x}_{i},
	\label{eqSpinLGTModel}
\end{multline}
where the factor $(4\hat{S}^{x}_{i-1}\hat{S}^{x}_{i+1}-1)$ in the hopping term is necessary to obtain the same matrix elements as in the original model, namely the hopping is only allowed when the original lattice site is occupied by a hard-core boson and the prospective lattice site, onto which the boson can hop, is empty. All other configurations do not allow hopping. The density-density correlator, $\langle \delta \hat{n}(x) \delta \hat{n}(0) \rangle$ was implemented in a similar way using the Eq.~\eqref{eqNumberGaugeFields}.

In order to calculate the the charge gap (see Supplementary E) we had to subtract the chemical potential contribution to the overall energy by modifying the chemical potential term to $\mu \sum_{i} ((0.5-n^a)-2\hat{S}^{x}_{i+1} \hat{S}^{x}_{i})$.

\emph{Simulations in the string-length basis.--}
Although the considered \Zt LGT expressed in terms of the string length basis Eq.~\eqref{eqHtransfrm} is a purely local Hamiltonian, the absence of the usual Bose enhancement makes an accurate study of this model challenging. In particular bosonic systems would require to consider a maximum occupation number equal to the total number of bosons $N$. Apart from the case of very small systems, this rigorous choice would make untreatable the bosonic problem with a quasi-exact method like DMRG. In order to achieve an efficient description, the usual strategy when studying standard bosonic models is to perform a cutoff in the local Hilbert space. This turns out to be a totally safe and precise manner to approach such systems especially in the regimes of low and intermediate densities and non attractive interactions. On the other hand, the peculiar structure of the string length model makes Hilbert space truncations very delicate. In particular we find that if we cut the maximum occupation number to too small values, but still large compared to the usual choices in "standard" bosonic models, we do not get agreement between the density-density correlation decay Eqs.~\eqref{eqNNcorrRel} and \eqref{DnstyCorrStrngLngth}. In order to achieve agreement and thus the correct Luttinger parameters relation Eq.~\eqref{eqRelLuttingerK} we employ extensive DMRG simulations by considering a maximum occupation number of $N/2$. This clearly requires a large number of DMRG states, that we fix $=1400$, in order to keep the truncation error on the energy $<10^{-8}$.

\subsection{Gap Calculations}

\begin{figure}[t!]
\centering
\epsfig{file=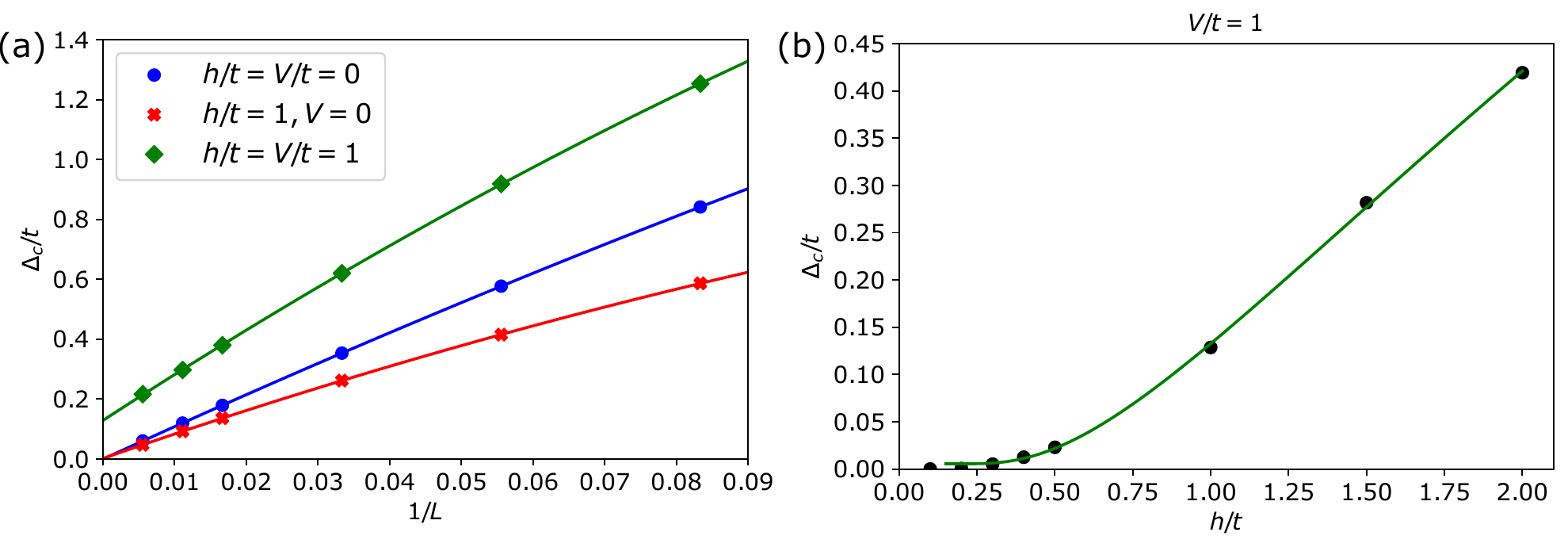, width=0.48\textwidth}
\caption{Gap extrapolation to the thermodynamic limit (a) by fitting the data points with a quadratic function: $h = V = 0$ (blue), $h/t = 1,V = 0$ (red) and $h/t = V/t = 1$ (green). (b) The opening of the gap as a function of $h$ at $V/t = 1$ (black data points) and the corresponding fit with the function $f_{\Delta} = A_{\Delta} +C_{\Delta} e^{-B_{\Delta}/\sqrt{h-h_{c}}}$. The critical value was set manually to $h_{c} / t = 0.15$. The filling was set to $n^a = 2/3$.}
\label{figGapExtrapAndBKT}
\end{figure}

In order to calculate the opening of the Mott gap in the thermodynamic limit the ground state energies were calculated for different chain lengths $L$. The energies were then plotted as a function of inverse chain length and fitted with a quadratic function, see Fig.~\ref{figGapExtrapAndBKT} (a). The value of the fitted equation at $x= 1/L =0$ was then taken as the value of ground state in the thermodynamic limit, $L \rightarrow \infty$. The filling was varied by tuning the chemical potential $\mu$ to obtained the correct value of $N$. By changing the chemical potential we also calculated the ground state for $N-2$ and $N+2$ in order to calculate the gap, $\Delta_{c}(L, N)$ using Eq.~\eqref{eqGap}. We furthermore note that the chemical potential contribution to the overall energy was deducted. By varying the parameters $h$ and $V$ the diagram in Fig.~\ref{figGapBars} was obtained.

The opening of the gap appears to behave similarly as in the BKT case, as is demonstrated in Fig.~\ref{figGapExtrapAndBKT} (b). There, the transition point was estimated to be $h_{c} / t \sim 0.15$ in order to produce a fit. Although the exact transition point is hard to deduce, the general behaviour appears to agree with an exponential gap opening.

\subsection{Particle-hole mapping}
The particle-hole mapping is performed by using the unitary transformation \cite{Prosko2017}
\begin{align}
	\ad_i & \rightarrow \a_i, \nonumber \\
	\a_i & \rightarrow \ad_i,
\label{eqPhUnitary}
\end{align}
and the Hamiltonian \eqref{eqDefModel} becomes
\begin{multline}
	\H = -t \sum_{\ij} \l \a_i \hat{\tau}^z_{\ij} \ad_j + \hc \r - h \sum_{\ij} \hat{\tau}^x_{\ij} \\
	+ V \sum_\ij (1-\hat{n}_i) (1-\hat{n}_j),
	\label{eqPhChargeMapHamiltonian}
\end{multline}
where the number operator was transformed by using the Eq.~\eqref{eqPhUnitary} 
\begin{equation}
	\n_{i} = \ad_{i}\a _{i}\rightarrow \a_{i}\ad_{i} = 1-\n_{i}.
	\label{eqDensityPhMap}
\end{equation}
In the last step the commutation relations of the hard-core bosons were used. Apart from the NN interaction term, Eq.~\eqref{eqPhChargeMapHamiltonian} has the exact same form as the original model \eqref{eqDefModel}. 

Due to Eq.~\eqref{eqDensityPhMap} the Gauss law maps to
\begin{multline}
	\hat{G}_{i} = \hat{\tau}^{x}_{\langle i-1,i \rangle} \hat{\tau}^{x}_{\langle i,i+1 \rangle} (-1)^{\n_{i}} \\
	\rightarrow \hat{\tau}^{x}_{\langle i-1,i \rangle} \hat{\tau}^{x}_{\langle i,i+1 \rangle} (-1)^{1- \n_{i}} = - \hat{G}_{i}.
\end{multline}
We can obtain the original form by performing unitary transformation $ \hat{\tau}^x_{\left \langle i, i+1 \right \rangle} \rightarrow (-1)^{i} \hat{\tau}^x_{\left \langle i, i+1 \right \rangle}$ and  $ \hat{\tau}^y_{\left \langle i, i+1 \right \rangle} \rightarrow (-1)^{i} \hat{\tau}^y_{\left \langle i, i+1 \right \rangle}$ but $ \hat{\tau}^z_{\left \langle i, i+1 \right \rangle} \rightarrow \hat{\tau}^z_{\left \langle i, i+1 \right \rangle}$ which brings the Gauss law in the same form as in the original model
\begin{multline}
	- \hat{\tau}^{x}_{\langle i-1,i \rangle} \hat{\tau}^{x}_{\langle i,i+1 \rangle} (-1)^{\n_{i}} \\
	\rightarrow - (-1)^{i-1} \hat{\tau}^{x}_{\langle i-1,i \rangle} (-1)^{i} \hat{\tau}^{x}_{\langle i,i+1 \rangle} (-1)^{\n_{i}} \\
	= \hat{\tau}^{x}_{\langle i-1,i \rangle} \hat{\tau}^{x}_{\langle i,i+1 \rangle} (-1)^{\n_{i}}.
\end{multline}
This last transformation of the gauge fields brings the Hamiltonian to its final form
\begin{multline}
	\H = -t \sum_{\ij} \l \ad_i \hat{\tau}^z_{\ij} \a_j + \hc \r - h \sum_{i} (-1)^{i} \hat{\tau}^x_{\langle i,i+1 \rangle} \\
	+ V \sum_\ij (1-\hat{n}_i) (1-\hat{n}_j).
	\label{eqPhMappedHamiltonian}
\end{multline}
As can be seen in Eq.~\eqref{eqPhMappedHamiltonian} the hopping term is identical to the original model whereas the field term $h$ acquires a staggered term $(-1)^{i}$ and there is no linear dependence on the string length. The NN interaction transforms as expected for a hard-core boson case.

\subsection{$V \rightarrow \infty$ limit}

\begin{figure}[t!]
\centering
\epsfig{file=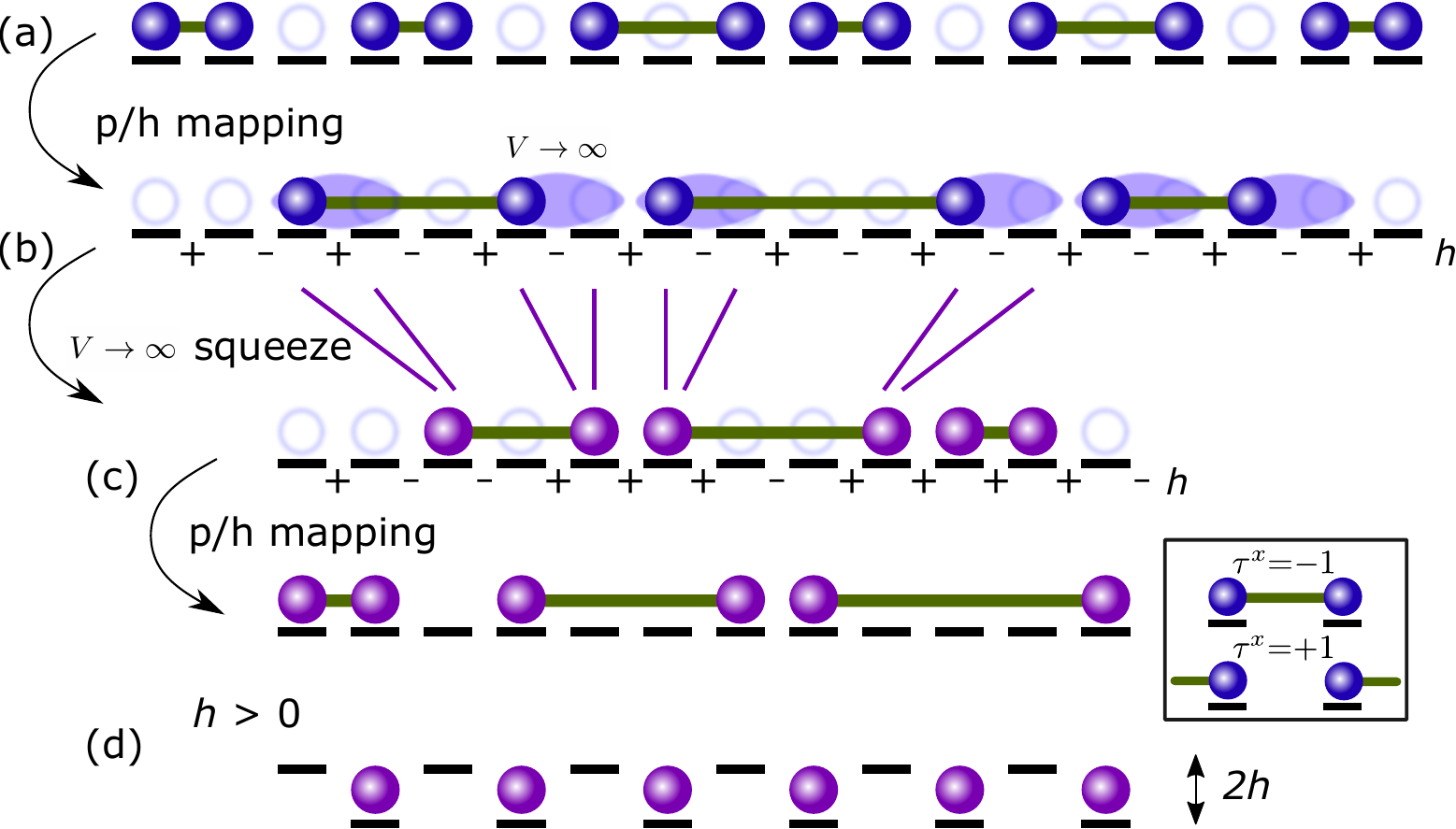, width=0.475\textwidth}
\caption{(a) Particle hole mapping. The $V \rightarrow \infty $ limit, which is taken in the next step, can be seen as a particle (depicted as a transparent oval) spreading to an empty site to the right of the occupied lattice site. The periodic exchange of $+$ and $-$ represent the staggered field $(-1)^{j} h$. (b) Mapping to a "quasi" squeezed space where the empty lattice sites following the occupied lattice sites are deleted from the picture and reintroduced in the Hamiltonian \eqref{eqVinfConstructionHamiltonianHole} as a new interaction term Eq.~\eqref{eqVinfConstructionFieldNewInteractions}. The shift of the staggered field due to relabelling of the lattice indices is again demonstrated with $+$ and $-$ bellow the lattice links. (c) Particle-hole mapping of the "quasi" squeezed space introduced above. (d) Expected ground state due to the staggered term in Hamiltonian \eqref{eqVinfHamiltonian} for half-filled lattice.}
\label{figVinfMappings}
\end{figure}

In this section we show that in the limit where $V \rightarrow \infty$ and $h > 0 $ we obtain at a gapped state for the filling of $n^a = 2/3$. We start by performing a particle-hole mapping of the original model as in the previous section, see Fig.~\ref{figVinfMappings} (a). In the particle-hole transformed picture the density becomes $n^{h} = 1/3$. Next we take the limit of $V \rightarrow \infty $ by making the particles larger and thus effectively occupying an extra lattice site to the right, see Fig.~\ref{figVinfMappings} (b). 
With such construction we make sure that there is at least one empty site  between two particles.

In the new picture the number of lattice sites decreases by $N$ and hence the density becomes $n' = \frac{N}{L-N} = \frac{n}{1-n}$ which for a filling $n^h = 1/3$ becomes $n' = 1/2$. Another important consequence of such construction is the fact that now one extra link variable and empty lattice site to the right of the occupied lattice site is hidden, see Fig.~\ref{figVinfMappings}. First of all, this means that the field in the new picture acquires an extra term
\begin{equation}
	-h \sum_{j=1}^{N} (-1)^{j} \hat{\tau}^{x}_{\langle j,j+1 \rangle} \rightarrow -h \sum_{j=1}^{L-N} (-1)^{j} \left( \prod_{i \leq j} (-1)^{\hat{n}_{i}} \right) \hat{\tau}^{x}_{\langle j,j+1 \rangle}
	\label{eqVinfConstructionField}
\end{equation}
where the site index $j$ in the new picture corresponds to $j + \sum_{i<j} \hat{n}_{i}$. Secondly, to account for the hidden link following each particle, we obtain a new interaction which can be written as
\begin{equation}
	-h \sum_{j=1}^{L-N} (-1)^{j} \left( \prod_{i<j} (-1)^{\hat{n}_{i}} \right) \hat{\tau}^{x}_{\langle j,j+1 \rangle} \hat{n}_{j}.
	\label{eqVinfConstructionFieldNewInteractions}
\end{equation}
The Hamiltonian in the new picture, where we took the limit $V \rightarrow \infty $, becomes
\begin{multline}
	\H '_{h} = -t \sum_{\ij} \l \ad_i \hat{\tau}^z_{\ij} \a_j + \hc \r \\ 
	-h \sum_{j} (-1)^{j} \left( \prod_{i \leq j} (-1)^{\hat{n}_{i}} \right) \hat{\tau}^{x}_{\langle i,i+1 \rangle} \\
	-h \sum_{j} (-1)^{j} \left( \prod_{i<j} (-1)^{\hat{n}_{i}} \right) \hat{\tau}^{x}_{\langle i,i+1 \rangle} \hat{n}_{j},
	\label{eqVinfConstructionHamiltonianHole}
\end{multline}
where the hopping term remained unchanged and we removed the $V$ term by incorporating an empty lattice site to the right of each particle. As a consequence we obtaine new field terms, which can be further simplified, while the Gauss law remains unchanged.

Before we simplify further we first perform another particle-hole transformation, this time in the new picture, Fig.~\ref{figVinfMappings} (c). Mathematically speaking we do the same thing as in the previous section where we replace $\hat{n}_{j} \rightarrow (1-\hat{n}_{j})$ and $\hat{\tau}^{x}_{\langle j,j+1 \rangle} \rightarrow (-1)^{j} \hat{\tau}^{x}_{\langle j,j+1 \rangle}$.

Eq.~\eqref{eqVinfConstructionField} becomes
\begin{multline}
	-h \sum_{j} (-1)^{j} \left( \prod_{i \leq j} (-1)^{\hat{n}_{i}} \right) \hat{\tau}^{x}_{\langle j,j+1 \rangle} \rightarrow \\
	\rightarrow -h \sum_{j} (-1)^{j} \left( \prod_{i \leq j} (-1)^{1- \hat{n}_{i}} \right) (-1)^{j} \hat{\tau}^{x}_{\langle j,j+1 \rangle} = \\
	= -h \sum_{j} (-1)^{j} \left( \prod_{i \leq j} (-1)^{\hat{n}_{i}} \right) \hat{\tau}^{x}_{\langle j,j+1 \rangle},
\end{multline}
where we used $\left( \prod_{i \leq j} (-1)^{1-\hat{n}_{i}} \right) = (-1)^{j} \left( \prod_{i \leq j} (-1)^{\hat{n}_{i}} \right)$, which results in exactly the same term as before. Similar manipulations are performed in the second term, Eq.~\eqref{eqVinfConstructionFieldNewInteractions} where in contrast the product term is written as $\prod_{i<j} (-1)^{\hat{n}_{i}} = -(-1)^{j} \prod_{i < j} (-1)^{\hat{n}_{i}}$. The Hamiltonian thus becomes
\begin{multline}
	\H ' = -t \sum_{\ij} \l \ad_i \hat{\tau}^z_{\ij} \a_j + \hc \r \\ 
	-h \sum_{j} (-1)^{j} \left( \prod_{i \leq j} (-1)^{\hat{n}_{i}} \right) \hat{\tau}^{x}_{\langle j,j+1 \rangle} \\
	+h \sum_{j} (-1)^{j} \left( \prod_{i<j} (-1)^{\hat{n}_{i}} \right) \hat{\tau}^{x}_{\langle j,j+1 \rangle} \left (1- \hat{n}_{j} \right ).
	\label{eqVinfConstructionHamiltonian}
\end{multline}
Using the Gauss law to write $\prod_{i \leq j} (-1)^{\hat{n}_{i}} = \hat{\tau}^{x}_{\langle 0, 1 \rangle} \hat{\tau}^{x}_{\langle j,j+1 \rangle}$, and taking into account that $ \left (\hat{\tau}^{x}_{\langle j,j+1 \rangle} \right)^2 = 1$ and the boundary condition $\hat{\tau}^{x}_{\langle 0, 1 \rangle} = 1$, the field terms can now be simplified:
\begin{multline}
	-h \sum_{j} (-1)^{j} \left( \prod_{i \leq j} (-1)^{\hat{n}_{i}} \right) \hat{\tau}^{x}_{\langle j,j+1 \rangle} + \\
	+h \sum_{j} (-1)^{j} \left( \prod_{i<j} (-1)^{\hat{n}_{i}} \right) \hat{\tau}^{x}_{\langle j,j+1 \rangle} \left (1- \hat{n}_{j} \right ) = \\
	= -h \sum_{j} (-1)^{j} +h \sum_{j} (-1)^{j} (-1)^{\hat{n}_j} \left (1- \hat{n}_{j} \right ) =\\
	= -h \sum_{j} (-1)^{j} \hat{n}_{j}.
\end{multline}

With this simplification we can write the Hamiltonian \eqref{eqVinfConstructionHamiltonian} as
\begin{equation}
	\H ' = -t \sum_{\ij} \l \ad_i \hat{\tau}^z_{\ij} \a_j + \hc \r  -h \sum_{j} (-1)^{j} \hat{n}_{j}.
	\label{eqVinfHamiltonian}
\end{equation}

Followig \cite{Prosko2017} we can remove $\hat{\tau}^{z}$ and fermionize the hard-core bosons, $\hat{a}_{j} = \left ( \prod_{i<j} \hat{\tau}^{z}_{\left \langle i, i+1 \right \rangle} \left ( -1 \right)^{\hat{n}_i} \right) \hat{c}_{j}$, which yields the final Hamiltonian
\begin{equation}
	\H ' = -t \sum_{\ij} \l \cd_i \c_j + \hc \r  -h \sum_{j} (-1)^{j} \cd_j \c_j.
	\label{eqVinfHamiltonianFinal}
\end{equation}
At half filling, $n' = 1/2$, we obtain a band insulator with band gap $\Delta = 2h$, as depicted in Fig.~\ref{figVinfMappings} (d).


\begin{thebibliography}{54}%
\makeatletter
\providecommand \@ifxundefined [1]{%
 \@ifx{#1\undefined}
}%
\providecommand \@ifnum [1]{%
 \ifnum #1\expandafter \@firstoftwo
 \else \expandafter \@secondoftwo
 \fi
}%
\providecommand \@ifx [1]{%
 \ifx #1\expandafter \@firstoftwo
 \else \expandafter \@secondoftwo
 \fi
}%
\providecommand \natexlab [1]{#1}%
\providecommand \enquote  [1]{``#1''}%
\providecommand \bibnamefont  [1]{#1}%
\providecommand \bibfnamefont [1]{#1}%
\providecommand \citenamefont [1]{#1}%
\providecommand \href@noop [0]{\@secondoftwo}%
\providecommand \href [0]{\begingroup \@sanitize@url \@href}%
\providecommand \@href[1]{\@@startlink{#1}\@@href}%
\providecommand \@@href[1]{\endgroup#1\@@endlink}%
\providecommand \@sanitize@url [0]{\catcode `\\12\catcode `\$12\catcode
  `\&12\catcode `\#12\catcode `\^12\catcode `\_12\catcode `\%12\relax}%
\providecommand \@@startlink[1]{}%
\providecommand \@@endlink[0]{}%
\providecommand \url  [0]{\begingroup\@sanitize@url \@url }%
\providecommand \@url [1]{\endgroup\@href {#1}{\urlprefix }}%
\providecommand \urlprefix  [0]{URL }%
\providecommand \Eprint [0]{\href }%
\providecommand \doibase [0]{http://dx.doi.org/}%
\providecommand \selectlanguage [0]{\@gobble}%
\providecommand \bibinfo  [0]{\@secondoftwo}%
\providecommand \bibfield  [0]{\@secondoftwo}%
\providecommand \translation [1]{[#1]}%
\providecommand \BibitemOpen [0]{}%
\providecommand \bibitemStop [0]{}%
\providecommand \bibitemNoStop [0]{.\EOS\space}%
\providecommand \EOS [0]{\spacefactor3000\relax}%
\providecommand \BibitemShut  [1]{\csname bibitem#1\endcsname}%
\let\auto@bib@innerbib\@empty
\bibitem [{\citenamefont {Kogut}(1979)}]{Kogut1979}%
  \BibitemOpen
  \bibfield  {author} {\bibinfo {author} {\bibfnamefont {J.~B.}\ \bibnamefont
  {Kogut}},\ }\href {\doibase 10.1103/revmodphys.51.659} {\bibfield  {journal}
  {\bibinfo  {journal} {Reviews of Modern Physics}\ }\textbf {\bibinfo {volume}
  {51}},\ \bibinfo {pages} {659} (\bibinfo {year} {1979})}\BibitemShut
  {NoStop}%
\bibitem [{\citenamefont {Wilson}(1974)}]{Wilson1974}%
  \BibitemOpen
  \bibfield  {author} {\bibinfo {author} {\bibfnamefont {K.~G.}\ \bibnamefont
  {Wilson}},\ }\href {\doibase 10.1103/physrevd.10.2445} {\bibfield  {journal}
  {\bibinfo  {journal} {Physical Review D}\ }\textbf {\bibinfo {volume} {10}},\
  \bibinfo {pages} {2445} (\bibinfo {year} {1974})}\BibitemShut {NoStop}%
\bibitem [{\citenamefont {Wen}(2004)}]{Wen2004}%
  \BibitemOpen
  \bibfield  {author} {\bibinfo {author} {\bibfnamefont {X.-G.}\ \bibnamefont
  {Wen}},\ }\href@noop {} {\emph {\bibinfo {title} {Quantum field theory of
  many-body systems}}}\ (\bibinfo  {publisher} {Oxford University Press},\
  \bibinfo {year} {2004})\BibitemShut {NoStop}%
\bibitem [{\citenamefont {Levin}\ and\ \citenamefont {Wen}(2005)}]{Levin2005}%
  \BibitemOpen
  \bibfield  {author} {\bibinfo {author} {\bibfnamefont {M.}~\bibnamefont
  {Levin}}\ and\ \bibinfo {author} {\bibfnamefont {X.-G.}\ \bibnamefont
  {Wen}},\ }\href {\doibase 10.1103/revmodphys.77.871} {\bibfield  {journal}
  {\bibinfo  {journal} {Reviews of Modern Physics}\ }\textbf {\bibinfo {volume}
  {77}},\ \bibinfo {pages} {871} (\bibinfo {year} {2005})}\BibitemShut
  {NoStop}%
\bibitem [{\citenamefont {Lee}\ \emph {et~al.}(2006)\citenamefont {Lee},
  \citenamefont {Nagaosa},\ and\ \citenamefont {Wen}}]{Lee2006}%
  \BibitemOpen
  \bibfield  {author} {\bibinfo {author} {\bibfnamefont {P.~A.}\ \bibnamefont
  {Lee}}, \bibinfo {author} {\bibfnamefont {N.}~\bibnamefont {Nagaosa}}, \ and\
  \bibinfo {author} {\bibfnamefont {X.-G.}\ \bibnamefont {Wen}},\ }\href
  {\doibase 10.1103/revmodphys.78.17} {\bibfield  {journal} {\bibinfo
  {journal} {Reviews of Modern Physics}\ }\textbf {\bibinfo {volume} {78}},\
  \bibinfo {pages} {17} (\bibinfo {year} {2006})}\BibitemShut {NoStop}%
\bibitem [{\citenamefont {Kitaev}(2003)}]{Kitaev2003}%
  \BibitemOpen
  \bibfield  {author} {\bibinfo {author} {\bibfnamefont {A.}~\bibnamefont
  {Kitaev}},\ }\href {\doibase https://doi.org/10.1016/S0003-4916(02)00018-0}
  {\bibfield  {journal} {\bibinfo  {journal} {Annals of Physics}\ }\textbf
  {\bibinfo {volume} {303}},\ \bibinfo {pages} {2} (\bibinfo {year}
  {2003})}\BibitemShut {NoStop}%
\bibitem [{\citenamefont {Troyer}\ and\ \citenamefont
  {Wiese}(2005)}]{Troyer2005}%
  \BibitemOpen
  \bibfield  {author} {\bibinfo {author} {\bibfnamefont {M.}~\bibnamefont
  {Troyer}}\ and\ \bibinfo {author} {\bibfnamefont {U.-J.}\ \bibnamefont
  {Wiese}},\ }\href {\doibase 10.1103/physrevlett.94.170201} {\bibfield
  {journal} {\bibinfo  {journal} {Physical Review Letters}\ }\textbf {\bibinfo
  {volume} {94}},\ \bibinfo {pages} {170201} (\bibinfo {year}
  {2005})}\BibitemShut {NoStop}%
\bibitem [{\citenamefont {Alford}\ \emph {et~al.}(2008)\citenamefont {Alford},
  \citenamefont {Schmitt}, \citenamefont {Rajagopal},\ and\ \citenamefont
  {Schäfer}}]{Alford2008}%
  \BibitemOpen
  \bibfield  {author} {\bibinfo {author} {\bibfnamefont {M.~G.}\ \bibnamefont
  {Alford}}, \bibinfo {author} {\bibfnamefont {A.}~\bibnamefont {Schmitt}},
  \bibinfo {author} {\bibfnamefont {K.}~\bibnamefont {Rajagopal}}, \ and\
  \bibinfo {author} {\bibfnamefont {T.}~\bibnamefont {Schäfer}},\ }\href
  {\doibase 10.1103/revmodphys.80.1455} {\bibfield  {journal} {\bibinfo
  {journal} {Reviews of Modern Physics}\ }\textbf {\bibinfo {volume} {80}},\
  \bibinfo {pages} {1455} (\bibinfo {year} {2008})}\BibitemShut {NoStop}%
\bibitem [{\citenamefont {Magnifico}\ \emph {et~al.}(2020)\citenamefont
  {Magnifico}, \citenamefont {Felser}, \citenamefont {Silvi},\ and\
  \citenamefont {Montangero}}]{Magnifico2020}%
  \BibitemOpen
  \bibfield  {author} {\bibinfo {author} {\bibfnamefont {G.}~\bibnamefont
  {Magnifico}}, \bibinfo {author} {\bibfnamefont {T.}~\bibnamefont {Felser}},
  \bibinfo {author} {\bibfnamefont {P.}~\bibnamefont {Silvi}}, \ and\ \bibinfo
  {author} {\bibfnamefont {S.}~\bibnamefont {Montangero}},\ }\href@noop {} {\
  (\bibinfo {year} {2020})},\ \Eprint {http://arxiv.org/abs/2011.10658}
  {arXiv:2011.10658 [hep-lat]} \BibitemShut {NoStop}%
\bibitem [{\citenamefont {Kuno}\ \emph {et~al.}(2017)\citenamefont {Kuno},
  \citenamefont {Sakane}, \citenamefont {Kasamatsu}, \citenamefont {Ichinose},\
  and\ \citenamefont {Matsui}}]{Kuno2017}%
  \BibitemOpen
  \bibfield  {author} {\bibinfo {author} {\bibfnamefont {Y.}~\bibnamefont
  {Kuno}}, \bibinfo {author} {\bibfnamefont {S.}~\bibnamefont {Sakane}},
  \bibinfo {author} {\bibfnamefont {K.}~\bibnamefont {Kasamatsu}}, \bibinfo
  {author} {\bibfnamefont {I.}~\bibnamefont {Ichinose}}, \ and\ \bibinfo
  {author} {\bibfnamefont {T.}~\bibnamefont {Matsui}},\ }\href {\doibase
  10.1103/physrevd.95.094507} {\bibfield  {journal} {\bibinfo  {journal}
  {Physical Review D}\ }\textbf {\bibinfo {volume} {95}},\ \bibinfo {pages}
  {094507} (\bibinfo {year} {2017})}\BibitemShut {NoStop}%
\bibitem [{\citenamefont {Wiese}(2013)}]{Wiese2013}%
  \BibitemOpen
  \bibfield  {author} {\bibinfo {author} {\bibfnamefont {U.-J.}\ \bibnamefont
  {Wiese}},\ }\href {\doibase 10.1002/andp.201300104} {\bibfield  {journal}
  {\bibinfo  {journal} {Annalen der Physik}\ }\textbf {\bibinfo {volume}
  {525}},\ \bibinfo {pages} {777} (\bibinfo {year} {2013})}\BibitemShut
  {NoStop}%
\bibitem [{\citenamefont {Zohar}\ \emph {et~al.}(2013)\citenamefont {Zohar},
  \citenamefont {Cirac},\ and\ \citenamefont {Reznik}}]{Zohar2013}%
  \BibitemOpen
  \bibfield  {author} {\bibinfo {author} {\bibfnamefont {E.}~\bibnamefont
  {Zohar}}, \bibinfo {author} {\bibfnamefont {J.~I.}\ \bibnamefont {Cirac}}, \
  and\ \bibinfo {author} {\bibfnamefont {B.}~\bibnamefont {Reznik}},\ }\href
  {\doibase 10.1103/physreva.88.023617} {\bibfield  {journal} {\bibinfo
  {journal} {Physical Review A}\ }\textbf {\bibinfo {volume} {88}},\ \bibinfo
  {pages} {023617} (\bibinfo {year} {2013})}\BibitemShut {NoStop}%
\bibitem [{\citenamefont {Zohar}\ \emph {et~al.}(2015)\citenamefont {Zohar},
  \citenamefont {Cirac},\ and\ \citenamefont {Reznik}}]{Zohar2015}%
  \BibitemOpen
  \bibfield  {author} {\bibinfo {author} {\bibfnamefont {E.}~\bibnamefont
  {Zohar}}, \bibinfo {author} {\bibfnamefont {J.~I.}\ \bibnamefont {Cirac}}, \
  and\ \bibinfo {author} {\bibfnamefont {B.}~\bibnamefont {Reznik}},\ }\href
  {\doibase 10.1088/0034-4885/79/1/014401} {\bibfield  {journal} {\bibinfo
  {journal} {Reports on Progress in Physics}\ }\textbf {\bibinfo {volume}
  {79}},\ \bibinfo {pages} {014401} (\bibinfo {year} {2015})}\BibitemShut
  {NoStop}%
\bibitem [{\citenamefont {Bender}\ \emph {et~al.}(2018)\citenamefont {Bender},
  \citenamefont {Zohar}, \citenamefont {Farace},\ and\ \citenamefont
  {Cirac}}]{Bender2018}%
  \BibitemOpen
  \bibfield  {author} {\bibinfo {author} {\bibfnamefont {J.}~\bibnamefont
  {Bender}}, \bibinfo {author} {\bibfnamefont {E.}~\bibnamefont {Zohar}},
  \bibinfo {author} {\bibfnamefont {A.}~\bibnamefont {Farace}}, \ and\ \bibinfo
  {author} {\bibfnamefont {J.~I.}\ \bibnamefont {Cirac}},\ }\href {\doibase
  10.1088/1367-2630/aadb71} {\bibfield  {journal} {\bibinfo  {journal} {New
  Journal of Physics}\ }\textbf {\bibinfo {volume} {20}},\ \bibinfo {pages}
  {093001} (\bibinfo {year} {2018})}\BibitemShut {NoStop}%
\bibitem [{\citenamefont {Dalmonte}\ and\ \citenamefont
  {Montangero}(2016)}]{Dalmonte2016}%
  \BibitemOpen
  \bibfield  {author} {\bibinfo {author} {\bibfnamefont {M.}~\bibnamefont
  {Dalmonte}}\ and\ \bibinfo {author} {\bibfnamefont {S.}~\bibnamefont
  {Montangero}},\ }\href {\doibase 10.1080/00107514.2016.1151199} {\bibfield
  {journal} {\bibinfo  {journal} {Contemporary Physics}\ }\textbf {\bibinfo
  {volume} {57}},\ \bibinfo {pages} {388} (\bibinfo {year} {2016})}\BibitemShut
  {NoStop}%
\bibitem [{\citenamefont {Martinez}\ \emph {et~al.}(2016)\citenamefont
  {Martinez}, \citenamefont {Muschik}, \citenamefont {Schindler}, \citenamefont
  {Nigg}, \citenamefont {Erhard}, \citenamefont {Heyl}, \citenamefont {Hauke},
  \citenamefont {Dalmonte}, \citenamefont {Monz}, \citenamefont {Zoller},\ and\
  \citenamefont {Blatt}}]{Martinez2016}%
  \BibitemOpen
  \bibfield  {author} {\bibinfo {author} {\bibfnamefont {E.~A.}\ \bibnamefont
  {Martinez}}, \bibinfo {author} {\bibfnamefont {C.~A.}\ \bibnamefont
  {Muschik}}, \bibinfo {author} {\bibfnamefont {P.}~\bibnamefont {Schindler}},
  \bibinfo {author} {\bibfnamefont {D.}~\bibnamefont {Nigg}}, \bibinfo {author}
  {\bibfnamefont {A.}~\bibnamefont {Erhard}}, \bibinfo {author} {\bibfnamefont
  {M.}~\bibnamefont {Heyl}}, \bibinfo {author} {\bibfnamefont {P.}~\bibnamefont
  {Hauke}}, \bibinfo {author} {\bibfnamefont {M.}~\bibnamefont {Dalmonte}},
  \bibinfo {author} {\bibfnamefont {T.}~\bibnamefont {Monz}}, \bibinfo {author}
  {\bibfnamefont {P.}~\bibnamefont {Zoller}}, \ and\ \bibinfo {author}
  {\bibfnamefont {R.}~\bibnamefont {Blatt}},\ }\href {\doibase
  10.1038/nature18318} {\bibfield  {journal} {\bibinfo  {journal} {Nature}\
  }\textbf {\bibinfo {volume} {534}},\ \bibinfo {pages} {516} (\bibinfo {year}
  {2016})}\BibitemShut {NoStop}%
\bibitem [{\citenamefont {Schweizer}\ \emph {et~al.}(2019)\citenamefont
  {Schweizer}, \citenamefont {Grusdt}, \citenamefont {Berngruber},
  \citenamefont {Barbiero}, \citenamefont {Demler}, \citenamefont {Goldman},
  \citenamefont {Bloch},\ and\ \citenamefont {Aidelsburger}}]{Schweizer2019}%
  \BibitemOpen
  \bibfield  {author} {\bibinfo {author} {\bibfnamefont {C.}~\bibnamefont
  {Schweizer}}, \bibinfo {author} {\bibfnamefont {F.}~\bibnamefont {Grusdt}},
  \bibinfo {author} {\bibfnamefont {M.}~\bibnamefont {Berngruber}}, \bibinfo
  {author} {\bibfnamefont {L.}~\bibnamefont {Barbiero}}, \bibinfo {author}
  {\bibfnamefont {E.}~\bibnamefont {Demler}}, \bibinfo {author} {\bibfnamefont
  {N.}~\bibnamefont {Goldman}}, \bibinfo {author} {\bibfnamefont
  {I.}~\bibnamefont {Bloch}}, \ and\ \bibinfo {author} {\bibfnamefont
  {M.}~\bibnamefont {Aidelsburger}},\ }\href {\doibase
  10.1038/s41567-019-0649-7} {\bibfield  {journal} {\bibinfo  {journal} {Nature
  Physics}\ }\textbf {\bibinfo {volume} {15}},\ \bibinfo {pages} {1168}
  (\bibinfo {year} {2019})}\BibitemShut {NoStop}%
\bibitem [{\citenamefont {Görg}\ \emph {et~al.}(2019)\citenamefont {Görg},
  \citenamefont {Sandholzer}, \citenamefont {Minguzzi}, \citenamefont
  {Desbuquois}, \citenamefont {Messer},\ and\ \citenamefont
  {Esslinger}}]{Goerg2019}%
  \BibitemOpen
  \bibfield  {author} {\bibinfo {author} {\bibfnamefont {F.}~\bibnamefont
  {Görg}}, \bibinfo {author} {\bibfnamefont {K.}~\bibnamefont {Sandholzer}},
  \bibinfo {author} {\bibfnamefont {J.}~\bibnamefont {Minguzzi}}, \bibinfo
  {author} {\bibfnamefont {R.}~\bibnamefont {Desbuquois}}, \bibinfo {author}
  {\bibfnamefont {M.}~\bibnamefont {Messer}}, \ and\ \bibinfo {author}
  {\bibfnamefont {T.}~\bibnamefont {Esslinger}},\ }\href {\doibase
  10.1038/s41567-019-0615-4} {\bibfield  {journal} {\bibinfo  {journal} {Nature
  Physics}\ }\textbf {\bibinfo {volume} {15}},\ \bibinfo {pages} {1161}
  (\bibinfo {year} {2019})}\BibitemShut {NoStop}%
\bibitem [{\citenamefont {Mil}\ \emph {et~al.}(2020)\citenamefont {Mil},
  \citenamefont {Zache}, \citenamefont {Hegde}, \citenamefont {Xia},
  \citenamefont {Bhatt}, \citenamefont {Oberthaler}, \citenamefont {Hauke},
  \citenamefont {Berges},\ and\ \citenamefont {Jendrzejewski}}]{Mil2020}%
  \BibitemOpen
  \bibfield  {author} {\bibinfo {author} {\bibfnamefont {A.}~\bibnamefont
  {Mil}}, \bibinfo {author} {\bibfnamefont {T.~V.}\ \bibnamefont {Zache}},
  \bibinfo {author} {\bibfnamefont {A.}~\bibnamefont {Hegde}}, \bibinfo
  {author} {\bibfnamefont {A.}~\bibnamefont {Xia}}, \bibinfo {author}
  {\bibfnamefont {R.~P.}\ \bibnamefont {Bhatt}}, \bibinfo {author}
  {\bibfnamefont {M.~K.}\ \bibnamefont {Oberthaler}}, \bibinfo {author}
  {\bibfnamefont {P.}~\bibnamefont {Hauke}}, \bibinfo {author} {\bibfnamefont
  {J.}~\bibnamefont {Berges}}, \ and\ \bibinfo {author} {\bibfnamefont
  {F.}~\bibnamefont {Jendrzejewski}},\ }\href {\doibase
  10.1126/science.aaz5312} {\bibfield  {journal} {\bibinfo  {journal}
  {Science}\ }\textbf {\bibinfo {volume} {367}},\ \bibinfo {pages} {1128}
  (\bibinfo {year} {2020})}\BibitemShut {NoStop}%
\bibitem [{\citenamefont {Yang}\ \emph {et~al.}(2020)\citenamefont {Yang},
  \citenamefont {Sun}, \citenamefont {Ott}, \citenamefont {Wang}, \citenamefont
  {Zache}, \citenamefont {Halimeh}, \citenamefont {Yuan}, \citenamefont
  {Hauke},\ and\ \citenamefont {Pan}}]{Yang2020}%
  \BibitemOpen
  \bibfield  {author} {\bibinfo {author} {\bibfnamefont {B.}~\bibnamefont
  {Yang}}, \bibinfo {author} {\bibfnamefont {H.}~\bibnamefont {Sun}}, \bibinfo
  {author} {\bibfnamefont {R.}~\bibnamefont {Ott}}, \bibinfo {author}
  {\bibfnamefont {H.-Y.}\ \bibnamefont {Wang}}, \bibinfo {author}
  {\bibfnamefont {T.~V.}\ \bibnamefont {Zache}}, \bibinfo {author}
  {\bibfnamefont {J.~C.}\ \bibnamefont {Halimeh}}, \bibinfo {author}
  {\bibfnamefont {Z.-S.}\ \bibnamefont {Yuan}}, \bibinfo {author}
  {\bibfnamefont {P.}~\bibnamefont {Hauke}}, \ and\ \bibinfo {author}
  {\bibfnamefont {J.-W.}\ \bibnamefont {Pan}},\ }\href {\doibase
  10.1038/s41586-020-2910-8} {\bibfield  {journal} {\bibinfo  {journal}
  {Nature}\ }\textbf {\bibinfo {volume} {587}},\ \bibinfo {pages} {392}
  (\bibinfo {year} {2020})}\BibitemShut {NoStop}%
\bibitem [{\citenamefont {Zohar}\ \emph {et~al.}(2017)\citenamefont {Zohar},
  \citenamefont {Farace}, \citenamefont {Reznik},\ and\ \citenamefont
  {Cirac}}]{Zohar2017}%
  \BibitemOpen
  \bibfield  {author} {\bibinfo {author} {\bibfnamefont {E.}~\bibnamefont
  {Zohar}}, \bibinfo {author} {\bibfnamefont {A.}~\bibnamefont {Farace}},
  \bibinfo {author} {\bibfnamefont {B.}~\bibnamefont {Reznik}}, \ and\ \bibinfo
  {author} {\bibfnamefont {J.~I.}\ \bibnamefont {Cirac}},\ }\href {\doibase
  10.1103/physrevlett.118.070501} {\bibfield  {journal} {\bibinfo  {journal}
  {Physical Review Letters}\ }\textbf {\bibinfo {volume} {118}},\ \bibinfo
  {pages} {070501} (\bibinfo {year} {2017})}\BibitemShut {NoStop}%
\bibitem [{\citenamefont {Barbiero}\ \emph {et~al.}(2019)\citenamefont
  {Barbiero}, \citenamefont {Schweizer}, \citenamefont {Aidelsburger},
  \citenamefont {Demler}, \citenamefont {Goldman},\ and\ \citenamefont
  {Grusdt}}]{Barbiero2019}%
  \BibitemOpen
  \bibfield  {author} {\bibinfo {author} {\bibfnamefont {L.}~\bibnamefont
  {Barbiero}}, \bibinfo {author} {\bibfnamefont {C.}~\bibnamefont {Schweizer}},
  \bibinfo {author} {\bibfnamefont {M.}~\bibnamefont {Aidelsburger}}, \bibinfo
  {author} {\bibfnamefont {E.}~\bibnamefont {Demler}}, \bibinfo {author}
  {\bibfnamefont {N.}~\bibnamefont {Goldman}}, \ and\ \bibinfo {author}
  {\bibfnamefont {F.}~\bibnamefont {Grusdt}},\ }\href {\doibase
  10.1126/sciadv.aav7444} {\bibfield  {journal} {\bibinfo  {journal} {Science
  Advances}\ }\textbf {\bibinfo {volume} {5}} (\bibinfo {year} {2019}),\
  10.1126/sciadv.aav7444}\BibitemShut {NoStop}%
\bibitem [{\citenamefont {Homeier}\ \emph {et~al.}(2020)\citenamefont
  {Homeier}, \citenamefont {Schweizer}, \citenamefont {Aidelsburger},
  \citenamefont {Fedorov},\ and\ \citenamefont {Grusdt}}]{Homeier2020}%
  \BibitemOpen
  \bibfield  {author} {\bibinfo {author} {\bibfnamefont {L.}~\bibnamefont
  {Homeier}}, \bibinfo {author} {\bibfnamefont {C.}~\bibnamefont {Schweizer}},
  \bibinfo {author} {\bibfnamefont {M.}~\bibnamefont {Aidelsburger}}, \bibinfo
  {author} {\bibfnamefont {A.}~\bibnamefont {Fedorov}}, \ and\ \bibinfo
  {author} {\bibfnamefont {F.}~\bibnamefont {Grusdt}},\ }\href@noop {} {\
  (\bibinfo {year} {2020})},\ \Eprint {http://arxiv.org/abs/2012.05235}
  {arXiv:2012.05235 [quant-ph]} \BibitemShut {NoStop}%
\bibitem [{\citenamefont {Sedgewick}\ \emph {et~al.}(2002)\citenamefont
  {Sedgewick}, \citenamefont {Scalapino},\ and\ \citenamefont
  {Sugar}}]{Sedgewick2002}%
  \BibitemOpen
  \bibfield  {author} {\bibinfo {author} {\bibfnamefont {R.}~\bibnamefont
  {Sedgewick}}, \bibinfo {author} {\bibfnamefont {D.}~\bibnamefont
  {Scalapino}}, \ and\ \bibinfo {author} {\bibfnamefont {R.}~\bibnamefont
  {Sugar}},\ }\href {\doibase 10.1103/physrevb.65.054508} {\bibfield  {journal}
  {\bibinfo  {journal} {Physical Review B}\ }\textbf {\bibinfo {volume} {65}},\
  \bibinfo {pages} {054508} (\bibinfo {year} {2002})}\BibitemShut {NoStop}%
\bibitem [{\citenamefont {Demler}\ \emph {et~al.}(2002)\citenamefont {Demler},
  \citenamefont {Nayak}, \citenamefont {Kee}, \citenamefont {Kim},\ and\
  \citenamefont {Senthil}}]{Demler2002}%
  \BibitemOpen
  \bibfield  {author} {\bibinfo {author} {\bibfnamefont {E.}~\bibnamefont
  {Demler}}, \bibinfo {author} {\bibfnamefont {C.}~\bibnamefont {Nayak}},
  \bibinfo {author} {\bibfnamefont {H.-Y.}\ \bibnamefont {Kee}}, \bibinfo
  {author} {\bibfnamefont {Y.~B.}\ \bibnamefont {Kim}}, \ and\ \bibinfo
  {author} {\bibfnamefont {T.}~\bibnamefont {Senthil}},\ }\href {\doibase
  10.1103/physrevb.65.155103} {\bibfield  {journal} {\bibinfo  {journal}
  {Physical Review B}\ }\textbf {\bibinfo {volume} {65}},\ \bibinfo {pages}
  {155103} (\bibinfo {year} {2002})}\BibitemShut {NoStop}%
\bibitem [{\citenamefont {Kaul}\ \emph {et~al.}(2007)\citenamefont {Kaul},
  \citenamefont {Kim}, \citenamefont {Sachdev},\ and\ \citenamefont
  {Senthil}}]{Kaul2007}%
  \BibitemOpen
  \bibfield  {author} {\bibinfo {author} {\bibfnamefont {R.~K.}\ \bibnamefont
  {Kaul}}, \bibinfo {author} {\bibfnamefont {Y.~B.}\ \bibnamefont {Kim}},
  \bibinfo {author} {\bibfnamefont {S.}~\bibnamefont {Sachdev}}, \ and\
  \bibinfo {author} {\bibfnamefont {T.}~\bibnamefont {Senthil}},\ }\href
  {\doibase 10.1038/nphys790} {\bibfield  {journal} {\bibinfo  {journal}
  {Nature Physics}\ }\textbf {\bibinfo {volume} {4}},\ \bibinfo {pages} {28}
  (\bibinfo {year} {2007})}\BibitemShut {NoStop}%
\bibitem [{\citenamefont {Sachdev}\ and\ \citenamefont
  {Chowdhury}(2016)}]{Sachdev2016}%
  \BibitemOpen
  \bibfield  {author} {\bibinfo {author} {\bibfnamefont {S.}~\bibnamefont
  {Sachdev}}\ and\ \bibinfo {author} {\bibfnamefont {D.}~\bibnamefont
  {Chowdhury}},\ }\href {\doibase 10.1093/ptep/ptw110} {\bibfield  {journal}
  {\bibinfo  {journal} {Progress of Theoretical and Experimental Physics}\
  }\textbf {\bibinfo {volume} {2016}},\ \bibinfo {pages} {12C102} (\bibinfo
  {year} {2016})}\BibitemShut {NoStop}%
\bibitem [{\citenamefont {Senthil}\ and\ \citenamefont
  {Fisher}(2000)}]{Senthil2000}%
  \BibitemOpen
  \bibfield  {author} {\bibinfo {author} {\bibfnamefont {T.}~\bibnamefont
  {Senthil}}\ and\ \bibinfo {author} {\bibfnamefont {M.~P.~A.}\ \bibnamefont
  {Fisher}},\ }\href {\doibase 10.1103/physrevb.62.7850} {\bibfield  {journal}
  {\bibinfo  {journal} {Physical Review B}\ }\textbf {\bibinfo {volume} {62}},\
  \bibinfo {pages} {7850} (\bibinfo {year} {2000})}\BibitemShut {NoStop}%
\bibitem [{\citenamefont {Lee}(2007)}]{Lee2007}%
  \BibitemOpen
  \bibfield  {author} {\bibinfo {author} {\bibfnamefont {P.~A.}\ \bibnamefont
  {Lee}},\ }\href {\doibase 10.1088/0034-4885/71/1/012501} {\bibfield
  {journal} {\bibinfo  {journal} {Reports on Progress in Physics}\ }\textbf
  {\bibinfo {volume} {71}},\ \bibinfo {pages} {012501} (\bibinfo {year}
  {2007})}\BibitemShut {NoStop}%
\bibitem [{\citenamefont {Gazit}\ \emph {et~al.}(2017)\citenamefont {Gazit},
  \citenamefont {Randeria},\ and\ \citenamefont {Vishwanath}}]{Gazit2017}%
  \BibitemOpen
  \bibfield  {author} {\bibinfo {author} {\bibfnamefont {S.}~\bibnamefont
  {Gazit}}, \bibinfo {author} {\bibfnamefont {M.}~\bibnamefont {Randeria}}, \
  and\ \bibinfo {author} {\bibfnamefont {A.}~\bibnamefont {Vishwanath}},\
  }\href {\doibase 10.1038/nphys4028} {\bibfield  {journal} {\bibinfo
  {journal} {Nature Physics}\ }\textbf {\bibinfo {volume} {13}},\ \bibinfo
  {pages} {484} (\bibinfo {year} {2017})}\BibitemShut {NoStop}%
\bibitem [{\citenamefont {Borla}\ \emph
  {et~al.}(2020{\natexlab{a}})\citenamefont {Borla}, \citenamefont
  {Jeevanesan}, \citenamefont {Pollmann},\ and\ \citenamefont
  {Moroz}}]{Borla2021}%
  \BibitemOpen
  \bibfield  {author} {\bibinfo {author} {\bibfnamefont {U.}~\bibnamefont
  {Borla}}, \bibinfo {author} {\bibfnamefont {B.}~\bibnamefont {Jeevanesan}},
  \bibinfo {author} {\bibfnamefont {F.}~\bibnamefont {Pollmann}}, \ and\
  \bibinfo {author} {\bibfnamefont {S.}~\bibnamefont {Moroz}},\ }\href@noop {}
  {\  (\bibinfo {year} {2020}{\natexlab{a}})},\ \Eprint
  {http://arxiv.org/abs/2012.08543} {arXiv:2012.08543 [cond-mat.str-el]}
  \BibitemShut {NoStop}%
\bibitem [{\citenamefont {Borla}\ \emph
  {et~al.}(2020{\natexlab{b}})\citenamefont {Borla}, \citenamefont {Verresen},
  \citenamefont {Shah},\ and\ \citenamefont {Moroz}}]{Borla2020a}%
  \BibitemOpen
  \bibfield  {author} {\bibinfo {author} {\bibfnamefont {U.}~\bibnamefont
  {Borla}}, \bibinfo {author} {\bibfnamefont {R.}~\bibnamefont {Verresen}},
  \bibinfo {author} {\bibfnamefont {J.}~\bibnamefont {Shah}}, \ and\ \bibinfo
  {author} {\bibfnamefont {S.}~\bibnamefont {Moroz}},\ }\href@noop {} {\
  (\bibinfo {year} {2020}{\natexlab{b}})},\ \Eprint
  {http://arxiv.org/abs/2010.00607} {arXiv:2010.00607 [cond-mat.str-el]}
  \BibitemShut {NoStop}%
\bibitem [{\citenamefont {Seifert}\ \emph {et~al.}(2020)\citenamefont
  {Seifert}, \citenamefont {Dong}, \citenamefont {Chulliparambil},
  \citenamefont {Vojta}, \citenamefont {Tu},\ and\ \citenamefont
  {Janssen}}]{Seifert2020}%
  \BibitemOpen
  \bibfield  {author} {\bibinfo {author} {\bibfnamefont {U.~F.}\ \bibnamefont
  {Seifert}}, \bibinfo {author} {\bibfnamefont {X.-Y.}\ \bibnamefont {Dong}},
  \bibinfo {author} {\bibfnamefont {S.}~\bibnamefont {Chulliparambil}},
  \bibinfo {author} {\bibfnamefont {M.}~\bibnamefont {Vojta}}, \bibinfo
  {author} {\bibfnamefont {H.-H.}\ \bibnamefont {Tu}}, \ and\ \bibinfo {author}
  {\bibfnamefont {L.}~\bibnamefont {Janssen}},\ }\href {\doibase
  10.1103/physrevlett.125.257202} {\bibfield  {journal} {\bibinfo  {journal}
  {Physical Review Letters}\ }\textbf {\bibinfo {volume} {125}},\ \bibinfo
  {pages} {257202} (\bibinfo {year} {2020})}\BibitemShut {NoStop}%
\bibitem [{\citenamefont {Grusdt}\ and\ \citenamefont
  {Pollet}(2020)}]{Grusdt2020}%
  \BibitemOpen
  \bibfield  {author} {\bibinfo {author} {\bibfnamefont {F.}~\bibnamefont
  {Grusdt}}\ and\ \bibinfo {author} {\bibfnamefont {L.}~\bibnamefont
  {Pollet}},\ }\href {\doibase 10.1103/physrevlett.125.256401} {\bibfield
  {journal} {\bibinfo  {journal} {Physical Review Letters}\ }\textbf {\bibinfo
  {volume} {125}},\ \bibinfo {pages} {256401} (\bibinfo {year}
  {2020})}\BibitemShut {NoStop}%
\bibitem [{\citenamefont {Gonz\'alez-Cuadra}\ \emph {et~al.}(2020)\citenamefont
  {Gonz\'alez-Cuadra}, \citenamefont {Tagliacozzo}, \citenamefont
  {Lewenstein},\ and\ \citenamefont {Bermudez}}]{Cuadra2020}%
  \BibitemOpen
  \bibfield  {author} {\bibinfo {author} {\bibfnamefont {D.}~\bibnamefont
  {Gonz\'alez-Cuadra}}, \bibinfo {author} {\bibfnamefont {L.}~\bibnamefont
  {Tagliacozzo}}, \bibinfo {author} {\bibfnamefont {M.}~\bibnamefont
  {Lewenstein}}, \ and\ \bibinfo {author} {\bibfnamefont {A.}~\bibnamefont
  {Bermudez}},\ }\href {\doibase 10.1103/PhysRevX.10.041007} {\bibfield
  {journal} {\bibinfo  {journal} {Phys. Rev. X}\ }\textbf {\bibinfo {volume}
  {10}},\ \bibinfo {pages} {041007} (\bibinfo {year} {2020})}\BibitemShut
  {NoStop}%
\bibitem [{\citenamefont {Borla}\ \emph
  {et~al.}(2020{\natexlab{c}})\citenamefont {Borla}, \citenamefont {Verresen},
  \citenamefont {Grusdt},\ and\ \citenamefont {Moroz}}]{Borla2020}%
  \BibitemOpen
  \bibfield  {author} {\bibinfo {author} {\bibfnamefont {U.}~\bibnamefont
  {Borla}}, \bibinfo {author} {\bibfnamefont {R.}~\bibnamefont {Verresen}},
  \bibinfo {author} {\bibfnamefont {F.}~\bibnamefont {Grusdt}}, \ and\ \bibinfo
  {author} {\bibfnamefont {S.}~\bibnamefont {Moroz}},\ }\href {\doibase
  10.1103/physrevlett.124.120503} {\bibfield  {journal} {\bibinfo  {journal}
  {Physical Review Letters}\ }\textbf {\bibinfo {volume} {124}},\ \bibinfo
  {pages} {120503} (\bibinfo {year} {2020}{\natexlab{c}})}\BibitemShut
  {NoStop}%
\bibitem [{\citenamefont {White}(1992)}]{White1992}%
  \BibitemOpen
  \bibfield  {author} {\bibinfo {author} {\bibfnamefont {S.~R.}\ \bibnamefont
  {White}},\ }\href {\doibase 10.1103/physrevlett.69.2863} {\bibfield
  {journal} {\bibinfo  {journal} {Physical Review Letters}\ }\textbf {\bibinfo
  {volume} {69}},\ \bibinfo {pages} {2863} (\bibinfo {year}
  {1992})}\BibitemShut {NoStop}%
\bibitem [{\citenamefont {Schollwöck}(2011)}]{Schollwoeck2011}%
  \BibitemOpen
  \bibfield  {author} {\bibinfo {author} {\bibfnamefont {U.}~\bibnamefont
  {Schollwöck}},\ }\href {\doibase 10.1016/j.aop.2010.09.012} {\bibfield
  {journal} {\bibinfo  {journal} {Annals of Physics}\ }\textbf {\bibinfo
  {volume} {326}},\ \bibinfo {pages} {96} (\bibinfo {year} {2011})}\BibitemShut
  {NoStop}%
\bibitem [{\citenamefont {Hubig}\ \emph {et~al.}()\citenamefont {Hubig},
  \citenamefont {Lachenmaier}, \citenamefont {Linden}, \citenamefont
  {Reinhard}, \citenamefont {Stenzel}, \citenamefont {Swoboda}, \citenamefont
  {Grundner},\ and\ \citenamefont {Mardazad}}]{hubig:_syten_toolk}%
  \BibitemOpen
  \bibfield  {author} {\bibinfo {author} {\bibfnamefont {C.}~\bibnamefont
  {Hubig}}, \bibinfo {author} {\bibfnamefont {F.}~\bibnamefont {Lachenmaier}},
  \bibinfo {author} {\bibfnamefont {N.-O.}\ \bibnamefont {Linden}}, \bibinfo
  {author} {\bibfnamefont {T.}~\bibnamefont {Reinhard}}, \bibinfo {author}
  {\bibfnamefont {L.}~\bibnamefont {Stenzel}}, \bibinfo {author} {\bibfnamefont
  {A.}~\bibnamefont {Swoboda}}, \bibinfo {author} {\bibfnamefont
  {M.}~\bibnamefont {Grundner}}, \ and\ \bibinfo {author} {\bibfnamefont
  {S.}~\bibnamefont {Mardazad}},\ }\href {https://syten.eu} {\enquote {\bibinfo
  {title} {The \textsc{SyTen} toolkit},}\ }\BibitemShut {NoStop}%
\bibitem [{\citenamefont {Hubig}(2017)}]{hubig17:_symmet_protec_tensor_networ}%
  \BibitemOpen
  \bibfield  {author} {\bibinfo {author} {\bibfnamefont {C.}~\bibnamefont
  {Hubig}},\ }\emph {\bibinfo {title} {Symmetry-Protected Tensor Networks}},\
  \href {https://edoc.ub.uni-muenchen.de/21348/} {Ph.D. thesis},\ \bibinfo
  {school} {LMU München} (\bibinfo {year} {2017})\BibitemShut {NoStop}%
\bibitem [{\citenamefont {Prosko}\ \emph {et~al.}(2017)\citenamefont {Prosko},
  \citenamefont {Lee},\ and\ \citenamefont {Maciejko}}]{Prosko2017}%
  \BibitemOpen
  \bibfield  {author} {\bibinfo {author} {\bibfnamefont {C.}~\bibnamefont
  {Prosko}}, \bibinfo {author} {\bibfnamefont {S.-P.}\ \bibnamefont {Lee}}, \
  and\ \bibinfo {author} {\bibfnamefont {J.}~\bibnamefont {Maciejko}},\ }\href
  {\doibase 10.1103/physrevb.96.205104} {\bibfield  {journal} {\bibinfo
  {journal} {Physical Review B}\ }\textbf {\bibinfo {volume} {96}},\ \bibinfo
  {pages} {205104} (\bibinfo {year} {2017})}\BibitemShut {NoStop}%
\bibitem [{\citenamefont {Gerbier}\ and\ \citenamefont
  {Dalibard}(2010)}]{Gerbier2010}%
  \BibitemOpen
  \bibfield  {author} {\bibinfo {author} {\bibfnamefont {F.}~\bibnamefont
  {Gerbier}}\ and\ \bibinfo {author} {\bibfnamefont {J.}~\bibnamefont
  {Dalibard}},\ }\href {\doibase 10.1088/1367-2630/12/3/033007} {\bibfield
  {journal} {\bibinfo  {journal} {New Journal of Physics}\ }\textbf {\bibinfo
  {volume} {12}},\ \bibinfo {pages} {033007} (\bibinfo {year}
  {2010})}\BibitemShut {NoStop}%
\bibitem [{\citenamefont {Yi}\ \emph {et~al.}(2008)\citenamefont {Yi},
  \citenamefont {Daley}, \citenamefont {Pupillo},\ and\ \citenamefont
  {Zoller}}]{Yi2008}%
  \BibitemOpen
  \bibfield  {author} {\bibinfo {author} {\bibfnamefont {W.}~\bibnamefont
  {Yi}}, \bibinfo {author} {\bibfnamefont {A.~J.}\ \bibnamefont {Daley}},
  \bibinfo {author} {\bibfnamefont {G.}~\bibnamefont {Pupillo}}, \ and\
  \bibinfo {author} {\bibfnamefont {P.}~\bibnamefont {Zoller}},\ }\href
  {\doibase 10.1088/1367-2630/10/7/073015} {\bibfield  {journal} {\bibinfo
  {journal} {New Journal of Physics}\ }\textbf {\bibinfo {volume} {10}},\
  \bibinfo {pages} {073015} (\bibinfo {year} {2008})}\BibitemShut {NoStop}%
\bibitem [{\citenamefont {Yang}\ \emph {et~al.}(2017)\citenamefont {Yang},
  \citenamefont {Dai}, \citenamefont {Sun}, \citenamefont {Reingruber},
  \citenamefont {Yuan},\ and\ \citenamefont {Pan}}]{Yang2017}%
  \BibitemOpen
  \bibfield  {author} {\bibinfo {author} {\bibfnamefont {B.}~\bibnamefont
  {Yang}}, \bibinfo {author} {\bibfnamefont {H.-N.}\ \bibnamefont {Dai}},
  \bibinfo {author} {\bibfnamefont {H.}~\bibnamefont {Sun}}, \bibinfo {author}
  {\bibfnamefont {A.}~\bibnamefont {Reingruber}}, \bibinfo {author}
  {\bibfnamefont {Z.-S.}\ \bibnamefont {Yuan}}, \ and\ \bibinfo {author}
  {\bibfnamefont {J.-W.}\ \bibnamefont {Pan}},\ }\href {\doibase
  10.1103/physreva.96.011602} {\bibfield  {journal} {\bibinfo  {journal}
  {Physical Review A}\ }\textbf {\bibinfo {volume} {96}},\ \bibinfo {pages}
  {011602} (\bibinfo {year} {2017})}\BibitemShut {NoStop}%
\bibitem{SMPhysRev}
See Supplemental Material at [\textit{URL will be inserted by publisher}] for details on the experimental realization of the $\mathbb{Z}_{2}$ LGT, the calculation of the relation between Luttinger parameters, the calculation of the Luttinger parameter in the confined phase, numerical calculations of the density-density correlations, the DMRG calculations, the determination of the charge gap, the particle-hole mapping and the $V\rightarrow \infty$ limit.

\bibitem [{\citenamefont {Batista}\ and\ \citenamefont
  {Ortiz}(2000)}]{Batista2000}%
  \BibitemOpen
\bibfield  {journal} {  }\bibfield  {author} {\bibinfo {author} {\bibfnamefont
  {C.~D.}\ \bibnamefont {Batista}}\ and\ \bibinfo {author} {\bibfnamefont
  {G.}~\bibnamefont {Ortiz}},\ }\href {\doibase 10.1103/physrevlett.85.4755}
  {\bibfield  {journal} {\bibinfo  {journal} {Physical Review Letters}\
  }\textbf {\bibinfo {volume} {85}},\ \bibinfo {pages} {4755} (\bibinfo {year}
  {2000})}\BibitemShut {NoStop}%
\bibitem [{\citenamefont {Montorsi}\ \emph {et~al.}(2020)\citenamefont
  {Montorsi}, \citenamefont {Fazzini},\ and\ \citenamefont
  {Barbiero}}]{Montorsi2020}%
  \BibitemOpen
  \bibfield  {author} {\bibinfo {author} {\bibfnamefont {A.}~\bibnamefont
  {Montorsi}}, \bibinfo {author} {\bibfnamefont {S.}~\bibnamefont {Fazzini}}, \
  and\ \bibinfo {author} {\bibfnamefont {L.}~\bibnamefont {Barbiero}},\ }\href
  {\doibase 10.1103/physreva.101.043618} {\bibfield  {journal} {\bibinfo
  {journal} {Physical Review A}\ }\textbf {\bibinfo {volume} {101}},\ \bibinfo
  {pages} {043618} (\bibinfo {year} {2020})}\BibitemShut {NoStop}%
\bibitem [{\citenamefont {Giamarchi}(2004)}]{Giamarchi2004}%
  \BibitemOpen
  \bibfield  {author} {\bibinfo {author} {\bibfnamefont {T.}~\bibnamefont
  {Giamarchi}},\ }\href
  {https://www.ebook.de/de/product/3263637/thierry_giamarchi_quantum_physics_in_one_dimension.html}
  {\emph {\bibinfo {title} {Quantum Physics in One Dimension}}}\ (\bibinfo
  {publisher} {Oxford University Press},\ \bibinfo {year} {2004})\BibitemShut
  {NoStop}%
\bibitem [{\citenamefont {Luther}\ and\ \citenamefont
  {Emery}(1974)}]{Luther1974}%
  \BibitemOpen
  \bibfield  {author} {\bibinfo {author} {\bibfnamefont {A.}~\bibnamefont
  {Luther}}\ and\ \bibinfo {author} {\bibfnamefont {V.~J.}\ \bibnamefont
  {Emery}},\ }\href {\doibase 10.1103/physrevlett.33.589} {\bibfield  {journal}
  {\bibinfo  {journal} {Physical Review Letters}\ }\textbf {\bibinfo {volume}
  {33}},\ \bibinfo {pages} {589} (\bibinfo {year} {1974})}\BibitemShut
  {NoStop}%
\bibitem [{\citenamefont {Saffman}\ \emph {et~al.}(2010)\citenamefont
  {Saffman}, \citenamefont {Walker},\ and\ \citenamefont
  {M{\o}lmer}}]{Saffman2010}%
  \BibitemOpen
  \bibfield  {author} {\bibinfo {author} {\bibfnamefont {M.}~\bibnamefont
  {Saffman}}, \bibinfo {author} {\bibfnamefont {T.~G.}\ \bibnamefont {Walker}},
  \ and\ \bibinfo {author} {\bibfnamefont {K.}~\bibnamefont {M{\o}lmer}},\
  }\href {\doibase 10.1103/revmodphys.82.2313} {\bibfield  {journal} {\bibinfo
  {journal} {Reviews of Modern Physics}\ }\textbf {\bibinfo {volume} {82}},\
  \bibinfo {pages} {2313} (\bibinfo {year} {2010})}\BibitemShut {NoStop}%
\bibitem [{\citenamefont {Henkel}\ \emph {et~al.}(2010)\citenamefont {Henkel},
  \citenamefont {Nath},\ and\ \citenamefont {Pohl}}]{Henkel2010}%
  \BibitemOpen
  \bibfield  {author} {\bibinfo {author} {\bibfnamefont {N.}~\bibnamefont
  {Henkel}}, \bibinfo {author} {\bibfnamefont {R.}~\bibnamefont {Nath}}, \ and\
  \bibinfo {author} {\bibfnamefont {T.}~\bibnamefont {Pohl}},\ }\href {\doibase
  10.1103/physrevlett.104.195302} {\bibfield  {journal} {\bibinfo  {journal}
  {Physical Review Letters}\ }\textbf {\bibinfo {volume} {104}},\ \bibinfo
  {pages} {195302} (\bibinfo {year} {2010})}\BibitemShut {NoStop}%
\bibitem [{\citenamefont {Ogata}\ and\ \citenamefont
  {Shiba}(1990)}]{Ogata1990}%
  \BibitemOpen
  \bibfield  {author} {\bibinfo {author} {\bibfnamefont {M.}~\bibnamefont
  {Ogata}}\ and\ \bibinfo {author} {\bibfnamefont {H.}~\bibnamefont {Shiba}},\
  }\href {\doibase 10.1103/physrevb.41.2326} {\bibfield  {journal} {\bibinfo
  {journal} {Physical Review B}\ }\textbf {\bibinfo {volume} {41}},\ \bibinfo
  {pages} {2326} (\bibinfo {year} {1990})}\BibitemShut {NoStop}%
\bibitem [{\citenamefont {Hilker}\ \emph {et~al.}(2017)\citenamefont {Hilker},
  \citenamefont {Salomon}, \citenamefont {Grusdt}, \citenamefont {Omran},
  \citenamefont {Boll}, \citenamefont {Demler}, \citenamefont {Bloch},\ and\
  \citenamefont {Gross}}]{Hilker2017}%
  \BibitemOpen
  \bibfield  {author} {\bibinfo {author} {\bibfnamefont {T.~A.}\ \bibnamefont
  {Hilker}}, \bibinfo {author} {\bibfnamefont {G.}~\bibnamefont {Salomon}},
  \bibinfo {author} {\bibfnamefont {F.}~\bibnamefont {Grusdt}}, \bibinfo
  {author} {\bibfnamefont {A.}~\bibnamefont {Omran}}, \bibinfo {author}
  {\bibfnamefont {M.}~\bibnamefont {Boll}}, \bibinfo {author} {\bibfnamefont
  {E.}~\bibnamefont {Demler}}, \bibinfo {author} {\bibfnamefont
  {I.}~\bibnamefont {Bloch}}, \ and\ \bibinfo {author} {\bibfnamefont
  {C.}~\bibnamefont {Gross}},\ }\href {\doibase 10.1126/science.aam8990}
  {\bibfield  {journal} {\bibinfo  {journal} {Science}\ }\textbf {\bibinfo
  {volume} {357}},\ \bibinfo {pages} {484} (\bibinfo {year}
  {2017})}\BibitemShut {NoStop}%
\bibitem [{\citenamefont {Auerbach}(1994)}]{Auerbach1994}%
  \BibitemOpen
  \bibfield  {author} {\bibinfo {author} {\bibfnamefont {A.}~\bibnamefont
  {Auerbach}},\ }\href@noop {} {\emph {\bibinfo {title} {Interacting electrons
  and quantum magnetism}}}\ (\bibinfo  {publisher} {Springer-Verlag},\ \bibinfo
  {year} {1994})\BibitemShut {NoStop}%
\end{thebibliography}
\end{document}